\documentclass[graphics,floatfix, footinbib,tightenlines,nobibnotes, aps, prb, twocolumn]{revtex4-1}
\usepackage{amsmath,amssymb,wasysym}
\usepackage{graphicx}% Include figure files
\usepackage{dcolumn}% Align table columns on decimal point
\usepackage{bm}% bold math
\usepackage{braket}
\usepackage{subcaption}
\usepackage{verbatim}
\usepackage{float}
\usepackage{bbold}
\usepackage{color}
\usepackage{xcolor}
\usepackage{relsize}
\usepackage{amsthm}
\usepackage{enumerate}
\usepackage{soul,xcolor}
\usepackage[T1]{fontenc}
\usepackage[colorlinks=true ,urlcolor=blue,urlbordercolor={0 1 1}]{hyperref}

\newcommand{\beq}{\begin{eqnarray}}
\newcommand{\eeq}{\end{eqnarray}}

\newcommand{\bsp}{\begin{split}}
\newcommand{\esp}{\end{split}}

\newcommand{\be}{\begin{equation}}
\newcommand{\ee}{\end{equation}}

\begin{document}

\setstcolor{red}

\title{SU(4) chiral spin liquid, exciton supersolid and electric detection in moir\'e bilayers}
\author{
Ya-Hui Zhang$^1$}
\author{
D. N. Sheng$^2$}
\author{
Ashvin Vishwanath$^1$}

\affiliation{$^1$Department of Physics, Harvard University, Cambridge, MA, USA
}
\affiliation{$^2$Department of Physics and Astronomy, California State University, Northridge, CA 91330
}

\date{\today}% It is always \today, today,

\begin{abstract}

We propose moir\'e bilayer as a platform where exotic quantum phases can be stabilized and electrically detected.  Moir\'e bilayer consists of two separate moir\'e superlattice layers coupled through the inter-layer Coulomb repulsion. In the small distance limit, an SU(4) spin can be formed by  combining layer pseudospin and the real spin.  As a concrete example, we study an SU(4) spin model on triangular lattice in the fundamental representation. By tuning a three-site ring exchange term $K\sim \frac{t^3}{U^2}$, we find SU(4) symmetric crystallized phase and an SU(4)$_1$ chiral spin liquid (CSL) at the balanced filling. We also predict two different exciton supersolid phases with inter-layer coherence at imbalanced filling under displacement field.  Especially, the system can simulate an SU(2) Bose-Einstein-condensation (BEC) by injecting inter-layer excitons into the  magnetically ordered Mott insulator at the layer polarized limit. Smoking gun evidences of these phases can be obtained by measuring the pseudo-spin transport in the counter-flow channel.

\end{abstract}

\pacs{Valid PACS appear here}% PACS, the Physics and Astronomy
\maketitle

\textbf{Introduction} It is now well appreciated that spin plays an important role in strongly correlated systems. In addition to simple ferromagnetic or anti-ferromagnetic ordered phases, electronic spins can form non-ordered phases such as spin liquids\cite{anderson1973resonating,anderson1987resonating,kalmeyer1987equivalence,balents2010spin,savary2016quantum,knolle2019field}.  Spin liquids have been found numerically in many spin $1/2$ lattice models\cite{yan2011,depenbrock2012,jiang2012,wang2013,iqbal2013,gong2015global,he2017signatures,zhu2018,hu2019dirac,gong2019chiral, PhysRevB.96.115115, zhu2020doped, szasz2021phase, wietek2021mott,chen2021quantum}, but there is still no well-established evidence in real experiments. One important reason is the difficulty of probing neutral spin excitation.   A direct probe of spin transport could provide  smoking gun evidence of certain spin liquids, such as spinon Fermi surface state and chiral spin liquid.   Alas,  measuring spin transport in traditional solid state systems is unfeasible.   Here, we propose to measure the transport of a pseudospin formed by the layer degree of freedom in an electronic material based on two Coulomb coupled moir\'e superlattices, which we dub as moir\'e bilayer.

\begin{figure}[ht]
\centering
\includegraphics[width=0.5\textwidth]{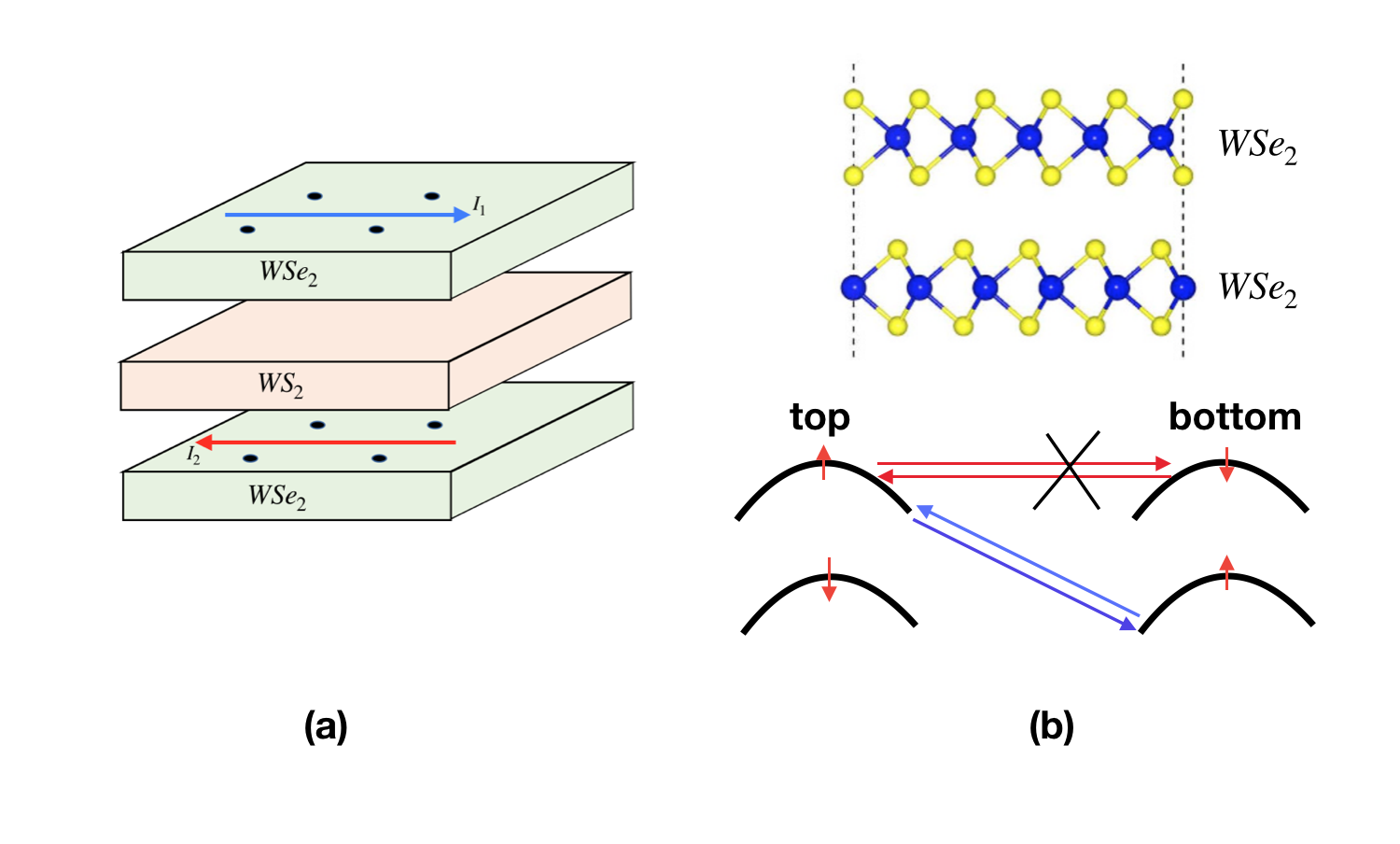}
\caption{Two ways of obtaining double moir\'e superlattice: (a) WSe$_2$-WS$_2$-WSe$_2$ sandwich with both WSe$_2$ layers aligned with WS$_2$. A triangular moir\'e superlattice can be generated for each WSe$_2$ due to the lattice mismatch between WSe$_2$ and WS$_2$\cite{wu2018hubbard,tang2020simulation,regan2019optical}. WS$_2$ also acts an insulating barrier to suppress inter-layer tunneling between the two WSe$_2$ layers.  (b) Twisted TMD homo-bilayer close to twist angle $\theta=60^\circ$. The top figure is a side view of bilayer WSe$_2$ system at angle $60^\circ$. The bottom figure illustrates the spin of the valence bands for the two TMD layers at the same valley, which leads to suppression of inter-layer tunneling  for the low energy moir\'e band generated at small twist angle.}
\label{fig:AB_TMD}
\end{figure}

To build a moir\'e bilayer, we wish to stack two 2D lattices and forbid their inter-layer tunneling. The total charge $N_a$ of each layer $a=1,2$ is separately conserved and we can label two quantum numbers as $Q=N_1+N_2$ and $P_z=\frac{1}{2}(N_1-N_2)$. $P_z$ can be viewed as a pseudo-spin. Actually, in the limit that the inter-layer distance $d$ is much smaller than the lattice constant $a_M$, there is a good SU(2) symmetry in the layer pseudospin space, similar to the well studied quantum Hall bilayer\cite{eisenstein2014exciton,li2017excitonic,liu2017quantum}.  Superlattices with $a_M \sim 10$ nm have been recently created  in several moir\'e systems based on graphene\cite{cao2018correlated,Wang2019Signatures,Wang2019Evidence,yankowitz2019tuning,Wang2019Signatures,chen2019tunable,lu2019superconductors,Cao2019Electric,Liu2019Spin,Shen2019observation,polshyn2020nonvolatile,chen2020electrically,sharpe2019emergent,serlin2020intrinsic} and transition metal dichalcogenides (TMD)\cite{tang2020simulation,regan2019optical,wang2019magic}. The moir\'e systems based on graphene generically exhibit ferromagnetic spin coupling due to band topology\cite{sharpe2019emergent,serlin2020intrinsic,Cao2019Electric,Shen2019observation,Liu2019Spin,chen2020electrically,polshyn2020nonvolatile,chen2019tunable}. To search for spin liquid, we will use moir\'e superlattice based on TMD as a building block, where anti-ferromagnetic spin coupling was demonstrated\cite{tang2020simulation}. We propose two different ways to generate  double moir\'e layers with two triangular moir\'e superlattices stacked together, as illustrated in Fig.~\ref{fig:AB_TMD}.

At integer total filling $\nu_T$, the system is in a Mott insulating phase if $U/t$ is large. There is a $SU(4)$ spin formed by the layer pseudospin $\vec P$ and the real spin $\vec S$.  Just as a concrete illustration, we focus on filling $\nu_T=1,3$ and map out the phase diagram of a $SU(4)$ spin model generated by $t/U$ expansion up to $O(\frac{t^3}{U^2})$.  One interesting phase we found is an $SU(4)_1$ chiral spin liquid stabilized by a three-site ring exchange term.   Chiral spin liquids\cite{kalmeyer1987equivalence,wen1989chiral} have been found to be the ground state for various spin 1/2 lattice models\cite{kagome_bauer_2014,he_2014,gong_2014,kagome_he_2015, zaletel_2020,hu_2016,wietek2015nature, Yao_2018,wietek2021mott,szasz2021phase,zhu2020doped,chen2021quantum,PhysRevB.96.115115} and also in $SU(N)$ model with $N>2$\cite{hermele2009mott,nataf2016chiral,Poilblanc_2020,boos2020time,yao2020topological,wu2016possible,tu2014quantum}. Compared to the early studies, the CSL in our model has a large spin gap (at order of J) and is stabilized in a wide range of $t/U$. More importantly, in the moir\'e bilayer setting up, smoking gun evidence of it can be obtained by measuring a quantized Hall effect of the layer pseudospin in counter-flow. Such electric probe of spin-Hall effect is impossible in previous proposals based on solid state spin and cold atom simulations.  In moir\'e bilayer, it is also easy to control the layer polarization $P_z$ continuously. When varying $P_z$ from $0$ to fully layer polarized, we also find two different supersolid phases with inter-layer coherence (exciton condensation) at small $P_z$ and large $P_z$ limit respectively. The imbalanced filling regime has not been explored  in previous studies of SU(N) model.

\textbf{Realization of SU(4) Hubbard model} We first derive an SU(4)  Hubbard model for moir\'e bilayer based on WSe$_2$-WS$_2$-WSe$_2$ or twisted AB stacked WSe$_2$ homo-bilayer, as illustrated in Fig.~\ref{fig:AB_TMD}. Both systems will host two triangular superlattices in the two WSe$_2$ layers.   In the supplementary we derive the lattice Hubbard model on triangular lattice by explicitly constructing Wannier orbitals and projecting the Coulomb interaction. One key ingredient is the suppression of the inter-layer tunneling due to either insulating barrier (WSe$_2$-WS$_2$-WSe$_2$) or spin conservation (twisted AB stacked WSe$_2$ bilayer). In the end we have four flavors by combining layer pseudospin and the real spin. The Low energy model is
\begin{equation}
 	H=-t \sum_{\langle ij \rangle}(c^\dagger_{i;\alpha}c_{j;\alpha}+h.c.)+\frac{U}{2} n_i(n_i-1)
 \end{equation} 
with $\alpha=a,\sigma$. $a=t,b$ is the pseudo-spin index which labels the top and bottom layer.  $\sigma=\uparrow,\downarrow$ labels the real spin (locked to the valley)\footnote{In TMD, the spin and valley are locked together due to a spin-orbit-coupling (SOC). We can view them together as a standard spin $1/2$ at zero magnetic field, but the g factor is anisotropic due to the SOC. Especially, the zeeman coupling to the out of plane magnetic field is large and the coupling to in-plane magnetic field is negligible.}.  $c^\dagger_{i;a,\sigma}$ creates an electron on moir\'e site $i$\footnote{Strictly speaking we are doping holes to the valence band of the TMD. But we will still call it "electron" to match the conventional language.}. There are also small easy-plane anisotropy terms due to  finite layer-separation. We will ignore them for now.

In this paper we will focus on the large $U/t$ regime at $\nu_T=1,3$, where there is a $SU(4)$ spin in the fundamental representation at each site. At filling $\nu_T=3$, at the large $U/t$ limit, the spin physics of the  Mott insulator is captured by the following $J-K$ model:

\begin{align}
H&=J \sum_{<ij>} P_{ij}+3 K \cos \Phi \sum_{<ijk> \in \bigtriangleup/\bigtriangledown}( P_{ijk}+P_{kji})\notag\\
&~~+3K \sin \Phi \sum_{<ijk> \in \bigtriangleup/\bigtriangledown}( i P_{ijk}-iP_{kji})
\end{align}
where each bond and each triangle should be counted only once.    $\Phi$ is the magnetic flux through each triangle. We will focus primarily on the $\Phi=0$ case with a time reversal symmetry.   We have $J=2 \frac{t^2}{U}-12\frac{t^3}{U^2}$ and $K=2\frac{t^3}{U^2}$.   For $\nu_T=1$, we just need to replace $t$ with $-t$.   In the above $P_{ij}$ and $P_{ijk}$ are two-site and three-site ring-exchange terms. For the triangular lattice, we define the two unit vectors to be $\mathbf a_1=(1,0)$ and $\mathbf a_2=(-\frac{1}{2},\frac{\sqrt{3}}{2})$. In the DMRG calculation, we use the boundary condition that $S(\mathbf r+L_y \mathbf a_2)=S(\mathbf r)$. The Hilbert space  at each site is constructed as a tensor product of two spin $1/2$ (layer pseudospin $\vec P$ and  real spin $\vec S$) and we label the corresponding Pauli matrix as $\tau_\mu$ and $\sigma_\mu$ respectively.  In this representation the generator of the $SU(4)$ can be labeled as $S_{\mu \nu}=\tau_\mu \otimes \sigma_\nu, \mu,\nu=0,x,y,z$.

\begin{figure}[ht]
\centering
\includegraphics[width=0.5\textwidth]{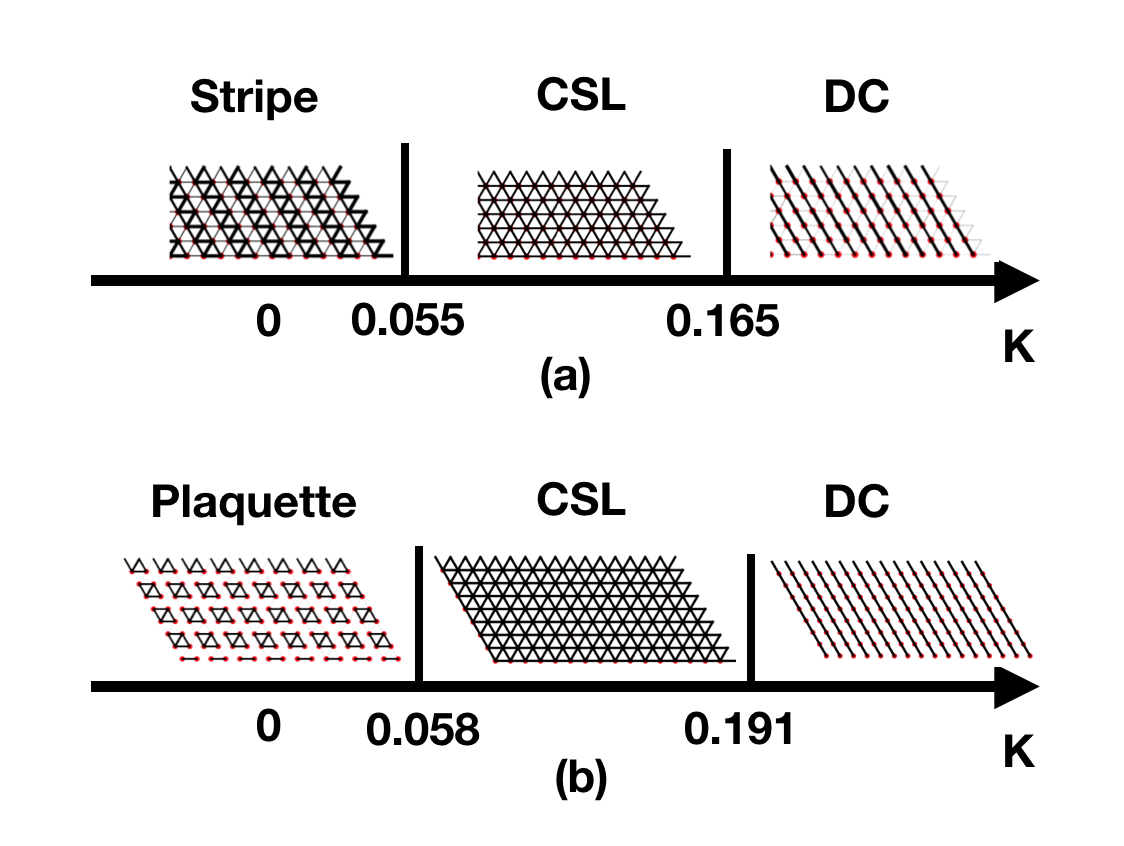}
\caption{Phase diagram from (a) DMRG and  (b) large-N mean field calculation (Note we have set $J=1$). CSL denotes the chiral spin liquid and DC denotes the decoupled chain phase.  In Fig.(a) we show typical patterns of bond order $\langle \tilde P_{ij}\rangle$  for the three phases. They are obtained for $K=0,\,K=0.114,\, {\rm and}\,K=0.27$ from finite DMRG calculation at $L_y=6$.  In DMRG calculation we find a stripe phase at $K=0$, but we believe it is unstable to plaquette order in the large $L_y$ limit (see the supplementary\cite{SM}). The phase boundaries in DMRG are based on $L_y=8$ and are already in fairly good agreement with the large-N result.}
\label{fig:phase_diagram}
\end{figure}

\textbf{Phase diagram at balanced filling:} We obtain a phase diagram at $\delta=0$ by tuning $K/J$ as shown in Fig.~\ref{fig:phase_diagram} by both DMRG simulation and large N mean field calculation. We find three phases: a crystal with $2\times 1$ or $2\times 2$ unit cell (spin crystal)\cite{yao2020topological}, a chiral spin liquid (CSL) and a phase with decoupled 1D chain (DC).   The CSL is in the range $K/J\in [0.055,0.165]$, or equivalently $t/U \in [0.041,0.082]$. At the upper critical value, higher order spin ring exchange terms may be needed\cite{yang2012effective}, which we leave to future work.  A remarkable observation is that the phase diagram obtained in DMRG  is qualitatively in  good agreement with that of a simple large $N$ mean field calculation, which suggests that $N=4$ may already be large enough to justify the mean field analysis.  Note that our result at the Heisenberg limit $K=0$ does not agree with a previous DMRG study\cite{keselman2020emergent} and we do not find signature of resonating plaquette order\cite{penc2003quantum}. For DMRG simulations, we keep the bond dimension to be between $4000-10000$ with a truncation error at the order of  $10^{-4}$ for $L_y=6$ and $8$ and smaller for $L_y=4$, providing  accurate results through finite bond dimension analysis (see Fig. 4 in the supplementary\cite{SM} for more details).

Let us also provide some intuition why the CSL and the DC phase are stabilized by $K>0$.   The three-site ring exchange term can be written as:
$
 \tilde P_{ijk}+h.c.=-8 [\vec S_i \cdot (\vec S_j \times \vec S_k)][\vec P_i \cdot (\vec P_j \times \vec P_k)]
 +2 \sum_{\tilde i \tilde j \tilde k}(\vec S_{\tilde i} \cdot \vec S_{\tilde k})(\vec P_{\tilde j} \cdot \vec P_{\tilde k})  
$.
When $K>0$, the first term favors onset of chirality order $\langle\vec S_i \cdot (\vec S_j \times \vec S_k)\rangle=\langle \vec P_i \cdot (\vec P_j \times \vec P_k) \rangle \neq 0$, leading to the CSL phase. The second term penalizes coexistence of two dimerized bonds for each triangle, favoring the decoupled chain phase. In contrast, the $K<0$ side suppresses chirality orders and favors plaquette order.

\begin{figure}[ht]
\centering
\includegraphics[width=0.5\textwidth]{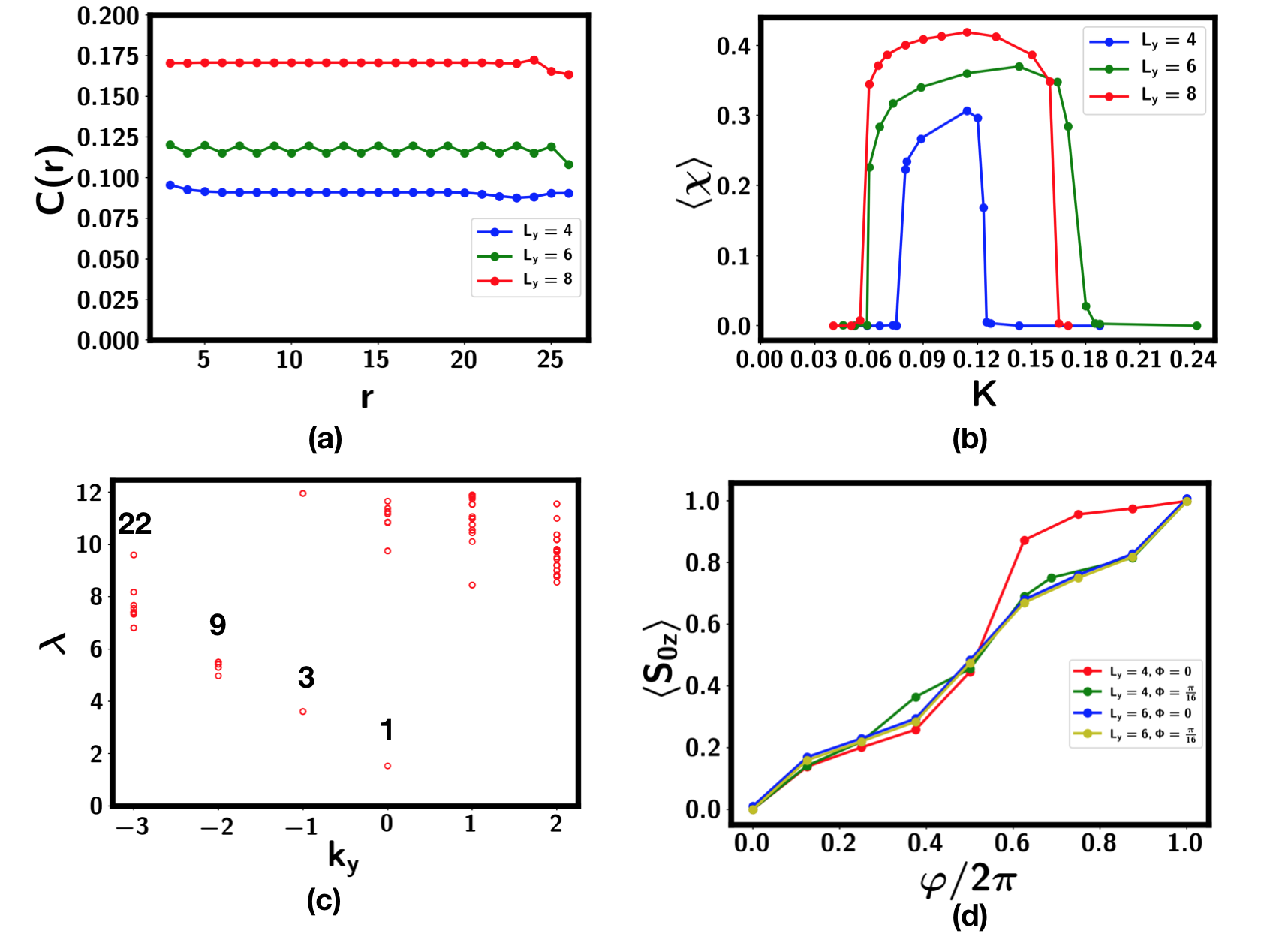}
\caption{(a) Correlation function of the chirality order at $K=0.114$ (we have set $J=1$ here) using a real code. $C(\mathbf r)=\langle \chi(\mathbf r) \chi(0)\rangle$, where $\chi=i(P_{ijk}-h.c.)$ is the chirality order parameter. (b) The chirality order $\langle \chi(r) \rangle$ obtained from finite DMRG with complex code, where $\chi_{ijk}=\langle i(P_{ijk}-h.c.)$ is defined for each triangle. It is clear that the CSL phase region is significantly enhanced when increasing $L_y$ from $4$ to $6$ or $8$.   (c) Entanglement spectrum from finite DMRG at $K=0.114$ and $\Phi=\frac{\pi}{16}$ for $L_y=6$. Weak explicit  time reversal breaking was included to enhance clarity. There is a chiral edge mode with degeneracy $1,3,9,22$. (d) Change in $\langle S_{0z} \rangle$ on the left side of the cylinder, pumped by the flux insertion generated by $U_1(\varphi)=e^{i \tilde Q_1 \varphi}$. Pumping of $S_{z0}$ and $S_{zz}$ are exactly the same and thus not shown. In the basis $ \tilde Q_1=\frac{1}{4}(S_{z0}+S_{0z}+S_{zz}), \tilde Q_2=\frac{1}{4}(S_{z0}-S_{0z}-S_{zz}), \tilde Q_3=\frac{1}{4}(-S_{z0}+S_{0z}-S_{zz})$, the pumped charges are $\delta \tilde Q_1=\frac{3}{4}$, $\delta \tilde Q_2=\delta \tilde Q_3=-\frac{1}{4}$.  }
\label{fig:csl_main}
\end{figure}

\textbf{The $SU(4)_1$ Chiral Spin Liquid:} Next we move to a detailed study of the CSL.  First, at $\Phi=0$, we find long range correlation of chirality order, as shown in Fig.~\ref{fig:csl_main}(a), suggesting spontaneous breaking of the time reversal symmetry. In Fig.~\ref{fig:csl_main}(b) we show the chirality order parameter with $K/J$ for $L_y=4,6,8$. We can see that the phase boundaries from $L_y=6$ and $L_y=8$ are close. In the supplementary we show that the CSL phase has a spin gap $\Delta_S \sim J$ and a correlation length $\xi_S<1$, therefore $L_y=6,8$ are much larger than the correlation length and may already be in the 2D limit. The $SU(4)_1$ CSL has a chiral edge described by the $SU(4)_1$ chiral CFT. It consists of three chiral boson and its entanglement spectrum should show a  degeneracy of $1,3,9,22,...$ for a given spin sector\cite{nataf2016chiral}.  Precisely such a  sequence is confirmed by our DMRG calculation in Fig.~\ref{fig:csl_main}(c).

 The CSL has a spin Hall conductivity $\sigma_{xy}$ which can be measured in DMRG via flux insertion\cite{Laughlin1981, gong_2014}.  For each quantum number $ \tilde Q_1=\frac{1}{4}(S_{z0}+S_{0z}+S_{zz}), \tilde Q_2=\frac{1}{4}(S_{z0}-S_{0z}-S_{zz}), \tilde Q_3=\frac{1}{4}(-S_{z0}+S_{0z}-S_{zz})$, we define  a twisted boundary condition $S(\mathbf r+Ly \mathbf{a_2})=U^\dagger_I(\varphi) S(\mathbf r) U_I(\varphi)$, where $U_I(\varphi)=e^{i \tilde Q_I \varphi}$ and $S(\mathbf r)$ is an arbitrary spin operator at site $\mathbf r$.    Note that $U_I(\varphi=2\pi)=e^{-i \frac{2\pi}{4}}I$ is a $Z_4$ flux insertion.   In Fig.~\ref{fig:csl_main}(b) we show the spin pumping generated by $U_1(\varphi)$, which implies spin Hall conductivity $\tilde \sigma_{xy}^{i1}=(\frac{3}{4},-\frac{1}{4},-\frac{1}{4})$ for $i=1,2,3$.  The pumping of $U_2(\varphi)$ and $U_3(\varphi)$ give consistent results and we get  $\tilde \sigma_{xy}= \frac14 \begin{pmatrix} +3 & -1 &-1 \\ -1 & +3 & -1 \\ -1 & -1 & +3 \end{pmatrix}$ which is nothing but the inverse of the K matrix: $K=\begin{pmatrix} 2 & 1 &1 \\ 1 &2 &1 \\ 1 &1 &2 \end{pmatrix}$\footnote{Note $\tilde \sigma_{xy}=K^{-1}$ holds only for the special basis of quantum numbers. For the simplest basis with $Q_1=S_{z0},Q_2=S_{0z},Q_3=S_{zz}$, we have $\sigma_{xy;ab}=2 \delta_{ab}$, with $a,b=1,2,3$. But the flux insertion generated by these simple quantum charges $Q_1,Q_2,Q_3$ is only $Z_2$, not the more fundamental $Z_4$ flux. To generate the $Z_4$ flux insertion, we need to use a complicated basis $\tilde Q_I$, which also lead to $\tilde \sigma_{xy}=K^{-1}$.}.

 We also studied the effect of SU(4) breaking anisotropy term $H_S=\delta J \sum_{\langle ij \rangle} (P_{i;x}P_{j;x}+P_{i;y}P_{j;y})(4\vec{S}_i\cdot \vec{S}_j+S_{i;0}S_{j;0})+2(\delta J+\delta V)\sum_{\langle ij \rangle}P_{i;z}P_{j;z}$ caused by the finite inter-layer distance.  We find that the CSL phase is stable when $\delta J/J<0.5, \delta V/J<0.5$ in DMRG calculation\cite{SM}, which is satisfied when the inter-layer distance $d< 1\, {\rm nm}$.

 \textbf{Supersolids at imbalanced filling} In the moir\'e bilayer setting up, we can also consider imbalanced filling with the density of the two layers to be $n_t=\frac{1}{2}+\frac{1}{2}\delta$ and $n_b=\frac{1}{2}-\frac{1}{2}\delta$.   We study the effect of non-zero $\delta$  by fixing $P_z=\frac{1}{2} \delta$ in the DMRG calculation (see Figure 4.) Here we note two supersolid phases found at $K=0$:

 \begin{figure}[ht]
\centering
\includegraphics[width=0.5\textwidth]{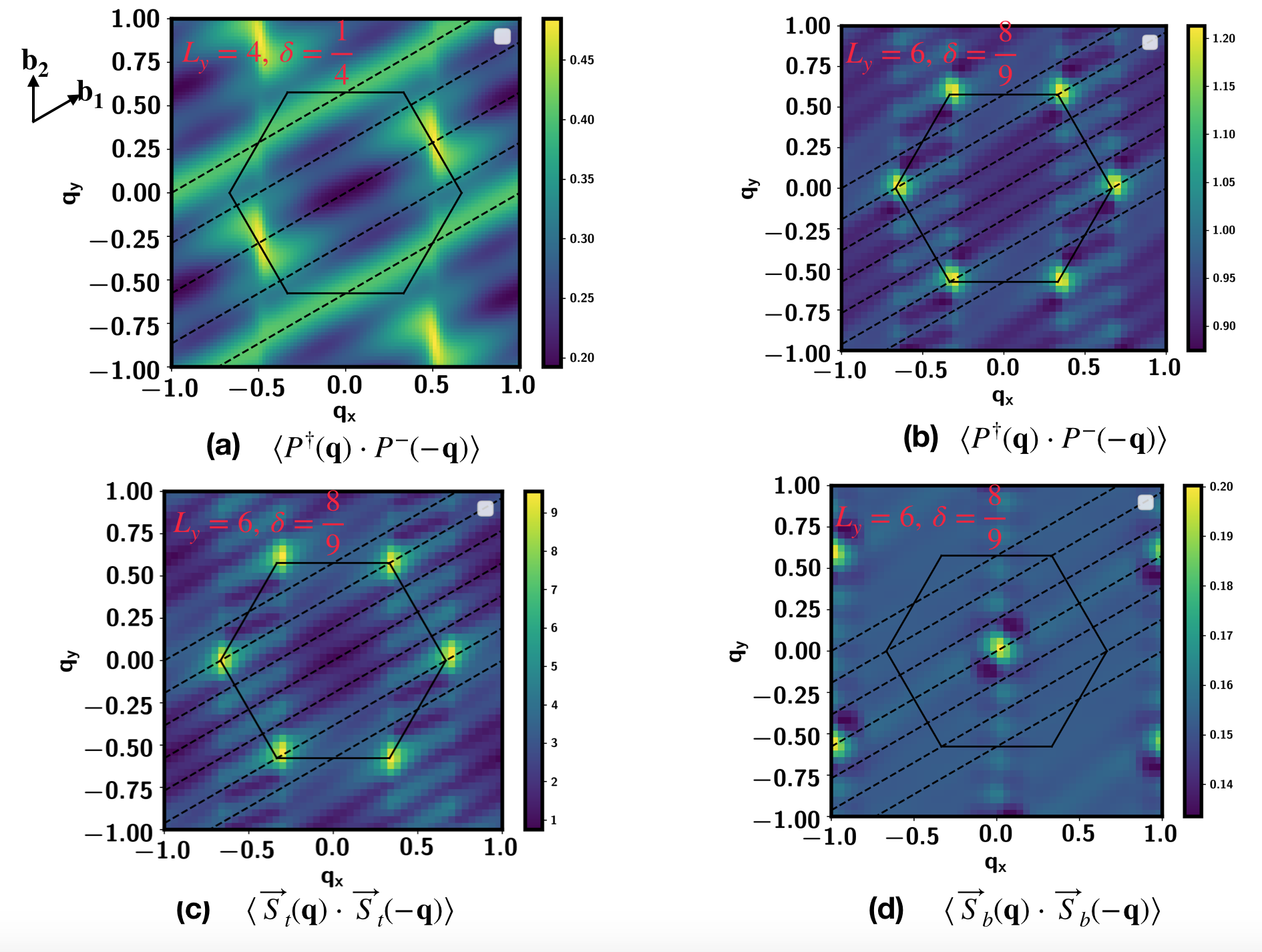}
\caption{Spin and exciton structure factor from Infinite DMRG at imbalanced filling. $q_x,q_y$ is in unit of $\frac{2\pi}{a}$, where $a$ is the lattice constant. We parameterize the momentum as $\mathbf{q}=q_1 \mathbf{b_1}+q_2 \mathbf{b_2}$, where $\mathbf{b_1}$ and $\mathbf{b_2}$ are reciprocal vectors. The solid hexagon is the Brillouin zone and the dashed lines are the well defined momentum cut along $q_2=\frac{1}{L_y} n$ with $n$ an integer. (a) Exciton order correlation function $P^\dagger(\mathbf q)P^{-}(-\mathbf q)$ has a peak at M point with momentum $\frac{1}{2} \mathbf{b_1}$ at small $\delta$. There is also feature along $q_2=\pm \frac{1}{2}$ cut without dispersion along $\mathbf{b_1}$. This is consistent with a decoupled stripe phase at the $\delta=0$ limit.  At $\delta$ close to 1: (b) the exciton order parameter is peaked at K,K' point. (c) The spin $\vec{S}_t$ at the top layer is ordered at momentum $K,K'$, consistent with a 120$^\circ$ order. (d) The spin $\vec{S}_b$ at the bottom layer is ferro-magnetically ordered.}
\label{fig:imbalanced}
\end{figure}

 \begin{itemize}
 \item \textbf{Supersolid on top of stripe phase}. When $\delta$ is small, DMRG shows a stripe phase with bond pattern similar to the $K=0$ point at $\delta=0$ in Fig.~\ref{fig:phase_diagram}(a).  On top of the stripe phase, we find  exciton condensation at momentum $M$, as indicated by  correlation function of exciton order $P^\dagger=P_x+iP_y$ shown in Fig.~\ref{fig:imbalanced}(a). The real spin in this phase is not ordered. The exciton condensate  has a spatial structure due to its non-zero momentum $M$ and hence can be called a supersolid phase,  see Fig. \ref{fig:imbalanced}(a).
 \item \textbf{Spinful BEC at the layer polarized limit}. When $\delta=1-2x$ with small $x$, we can start from the $120^\circ$ Neel order in the top layer at the $n_t=1,n_b=0$ limit and then inject inter-layer excitons with density $x$. The inter-layer exciton carries a SU(2) spin index from the bottom layer\footnote{If we start from the phase with top layer polarized and in a $120^\circ$ Neel order phase, the inter-layer exciton can be labeled as $\Phi_{i;\sigma}=c^\dagger_{i;b\sigma} c_{i;t}$, where the spin of $c_{i;t}$ is assumed to be along the $120^\circ$ order direction at site $i$. As a result, only the spin of the bottom layer enters the exciton. Note that our system has $SU(2)_t \times SU(2)_b$ spin rotation symmetry and $SU(2)_t$ is already broken by the $120^\circ$ order.}.  Finally the system simulates a gas of spinful bosons on triangular lattice at total density $x$. The ground state is known to be a spin polarized Bose-Einstein-condensation (BEC) of the excitons. The real spin in this phase is in the $120^\circ$ ordered and ferromagnetic ordered phase respectively for the two layers, as confirmed by DMRG results shown in Fig.~\ref{fig:imbalanced}(b)(c)(d). Two recent experiments studied the transferring of inter-layer excitons starting from a layer polarized Mott insulator\cite{gu2021dipolar,zhang2021correlated}.  The low energy physics of the exciton and spin in these systems should be very similar to the model we study here\cite{zhang2021fermionic_exciton}. Therefore our prediction of a spin $1/2$ BEC could be directly relevant to these experiments.
 \end{itemize} 

 \begin{figure}[ht]
\centering
\includegraphics[width=0.45\textwidth]{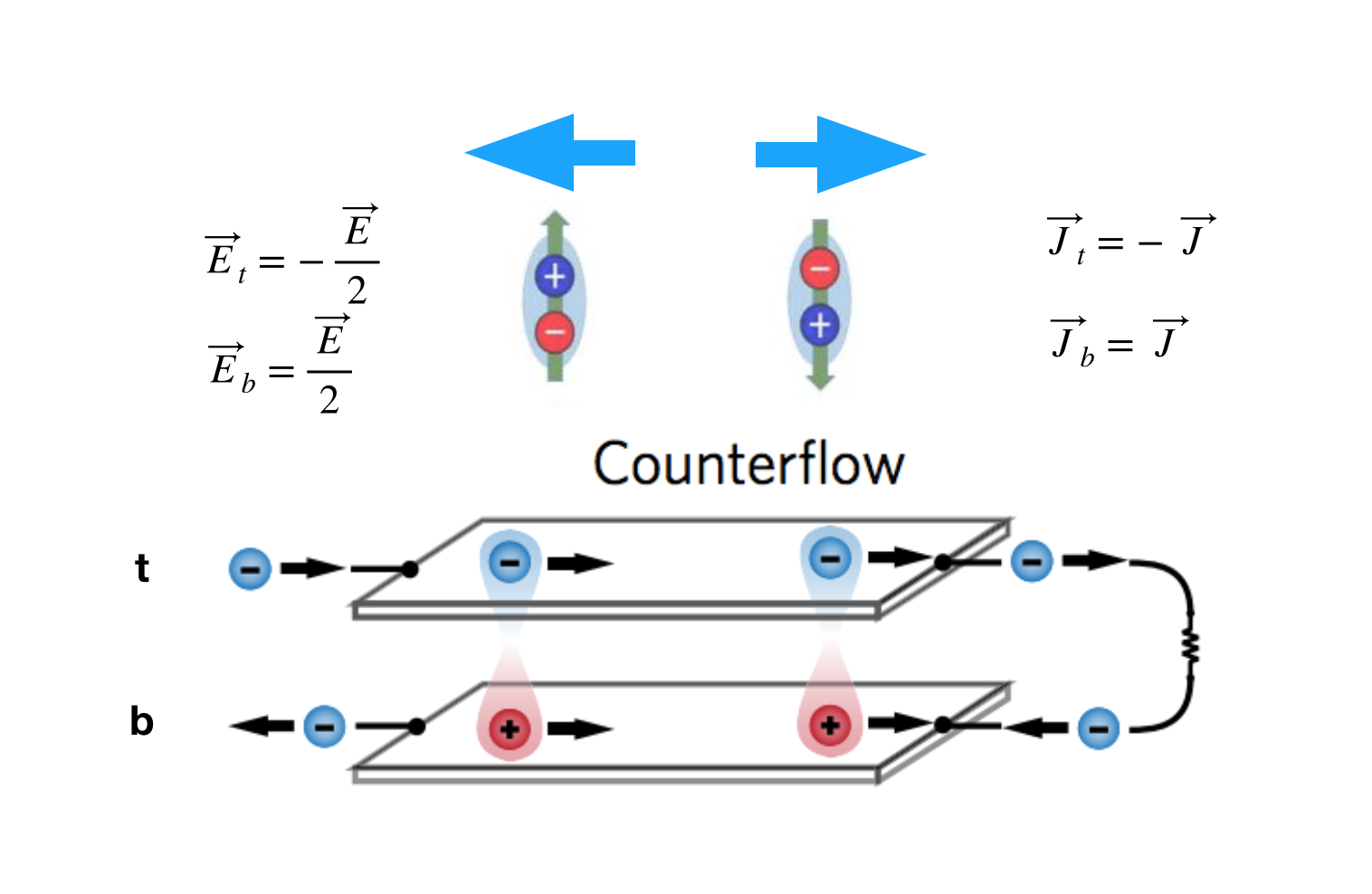}
\caption{Counterflow measurement of the transport of the electric dipole moment carried by the inter-layer exciton. $t,b$ labels the top and bottom layer respectively. $\vec E_d=\vec E_t-\vec E_b$ is the dipole electric field and $\vec J_d=\frac{1}{2} (\vec J_t-\vec J_b)$ is the dipole current. Under $\vec E_d$, the dipole moment feels a force $\vec F_d= P_z \vec E_d$. In the $SU(4)_1$ CSL, there is a dipole quantum Hall effect: $\sigma^{d}_{xy}=\frac{J^x_d}{E^y_d}=\pm \frac{e^2}{h}$. For the supersolid phase with inter-layer coherence, the counter-flow behavior is the same as a superfluid phase. }
\label{fig:counter_flow}
\end{figure}

\textbf{Experimental detection}  Here we point out that it is possible to obtain smoking gun evidences for the CSL phase and the supersolid phase in moir\'e bilayer  in counter-flow transport, as shown in Fig.~\ref{fig:counter_flow}. The counter-flow measures the current of the layer pseudo-spin $P_z$, which carries an electric dipole moment.  A dipole quantum Hall effect with $\sigma^{d}_{xy}=\pm \frac{e^2}{h}$ (see Fig.~\ref{fig:counter_flow}) is a direct evidence of the chiral spin liquid. For supersolid phase with inter-layer coherence, we expect a typical superfluid behavior with zero counter-flow resistivity.

\textbf{Summary} In conclusion, we proposed moir\'e bilayer as a new Hubbard model simulator, where the layer degree of freedom can simulate a pseudo-spin. This enables electric measurement of the pseudo-spin transport.  We focus on filling $\nu_T=1,3$ in the strong Mott limit, and find plaquette order, chiral spin liquid and supersolid phase. In the counter-flow transport, they will behave as trivial insulator, quantum Hall insulator and superfluid. In future, we hope to search for spinon Fermi surface and Dirac spin liquid in the Mott insulator, for which smoking gun evidence can also be obtained through metallic and semi-metallic pseudo-spin transport. We believe moir\'e bilayer is promising to shed light on strongly correlated problems with spin playing an essential role.

\textbf{Acknowledgement}  YHZ thanks T. Senthil for discussion and support for DMRG study at early stage of this work. AV and YHZ thank Cory Dean,  Philip Kim and Yihang Zeng for useful discussions.  AV and YHZ acknowledge support from a 2019 grant from the Harvard Quantum Initiative Seed Funding program, a Simons Investigator Fellowship  and the Simons Collaboration on Ultra-Quantum Matter, which is a grant from
the Simons Foundation (651440, A.V.). DNS was supported by the  U.S. Department of Energy, Office of Basic Energy Sciences under Grant No. DE-FG02-06ER46305.  The iDMRG simulation was performed using the
TeNPy Library (version 0.4.0)\cite{hauschild2018efficient}.

\bibliographystyle{apsrev4-1}
\bibliography{csl}

\appendix

\onecolumngrid

\section{Experimental realizations and lattice Hubbard model}

In this section we describe how to realize moir\'e bilayer in two different setting ups. The key idea is to have two Coulomb coupled layers with the same moir\'e superlattice. We need the inter-layer tunneling to be forbidden. In the twisted AB stacked TMD homo-bilayer, this is possible due to the spin conservation and opposite spin-valley locking in the two layers. In the WSe$_2$-WS$_2$-WSe$_2$ system, the middle WS$_2$ layer  provides moir\'e superlattice potential for the top and bottom WSe$_2$ layers and meanwhile also acts as an insulting barrier. 

\subsection{Twisted AB stacked TMD homo-bilayer}

\begin{figure}[ht]
\centering
\includegraphics[width=0.95\textwidth]{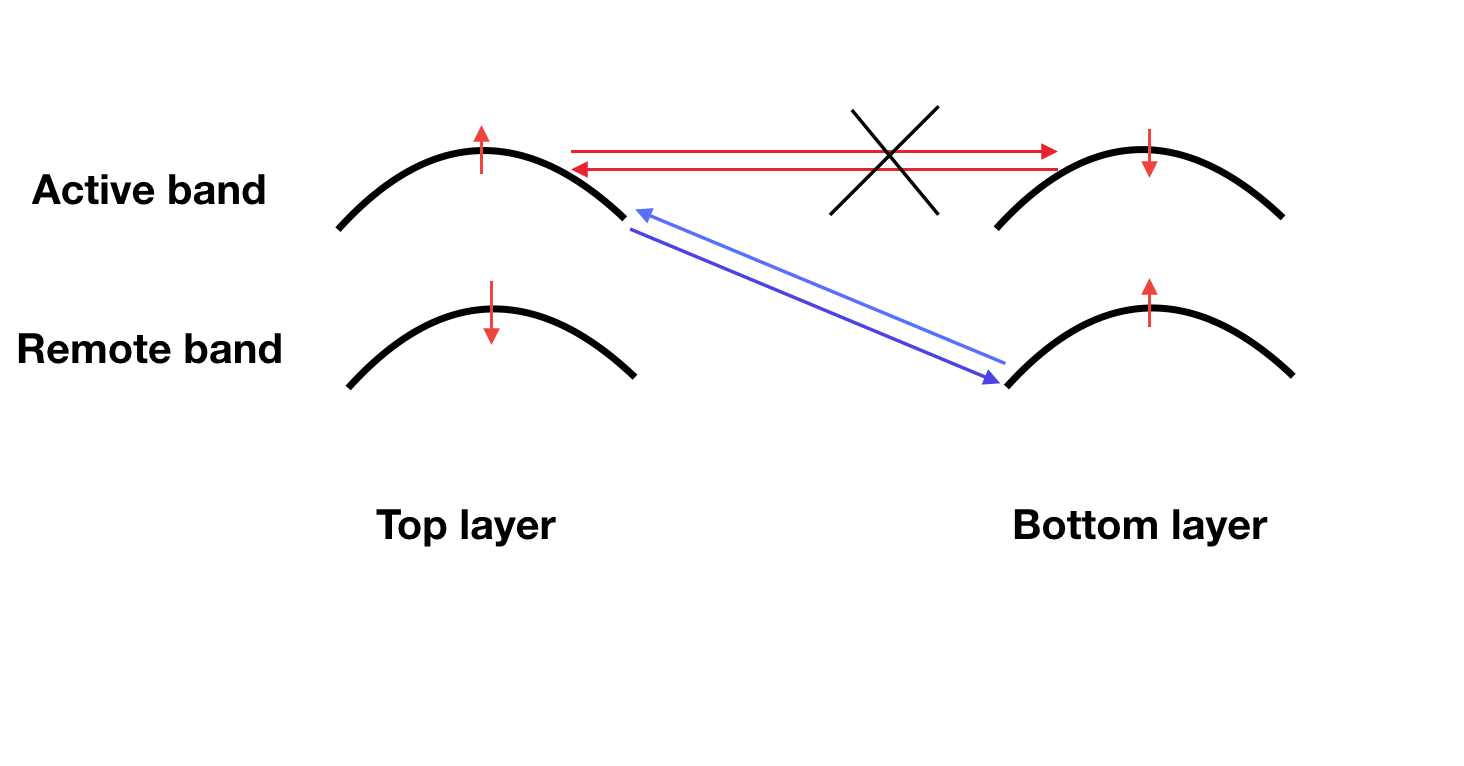}
\caption{Illustration of the top valence bands of the two layers at the same valley for the AB stacked TMD. The red arrow denotes the direction of the spin.  The inter-layer tunneling between the active bands are forbidden by the spin conservation. However, hole from the top layer can tunnel from the active band to the remote band of the bottom layer and then tunnels back. This second order process generates an intra-layer moir\'e superlattice potential.}
\label{fig:AB_TMD_appendix}
\end{figure}

We consider twisted TMD homobilayer at small twist angle $\theta$ starting from the AB stacking. In AB stacked TMD, one layer is rotated by $180^\circ$ compared to the AA stacked TMD. As a result, the definition of the two valley $K$ and $K'$ are interchanged just for one layer. Therefore we find that for the same valley, the two layers have opposite spin for the same band. Hence the inter-layer tunneling of the active band is forbidden. However, the inter-layer tunneling from the active band to the remote band is allowed and generates an intra-layer moir\'e superlattice potential through a second order process. We model the reconstructed band structure using the continuum model:

\begin{equation}
  H=H_0+H_M
\end{equation}

We have
\begin{equation}
  H_0= \sum_{\mathbf k} \frac{k^2}{2m^*}c^\dagger_{a\sigma}(\mathbf k) c_{a\sigma}(\mathbf k)
\end{equation}
where $a=t,b$ labels the two layers and $\sigma=\uparrow,\downarrow$ is the spin index. Note that the spin and the valley are locked in TMD, so the spin index is  also the valley index.  We use $m^*=0.62 m_e$ where $m_e$ is the bare electron mass\cite{wu2019topological}.

\begin{figure}[ht]
\centering
\includegraphics[width=0.95\textwidth]{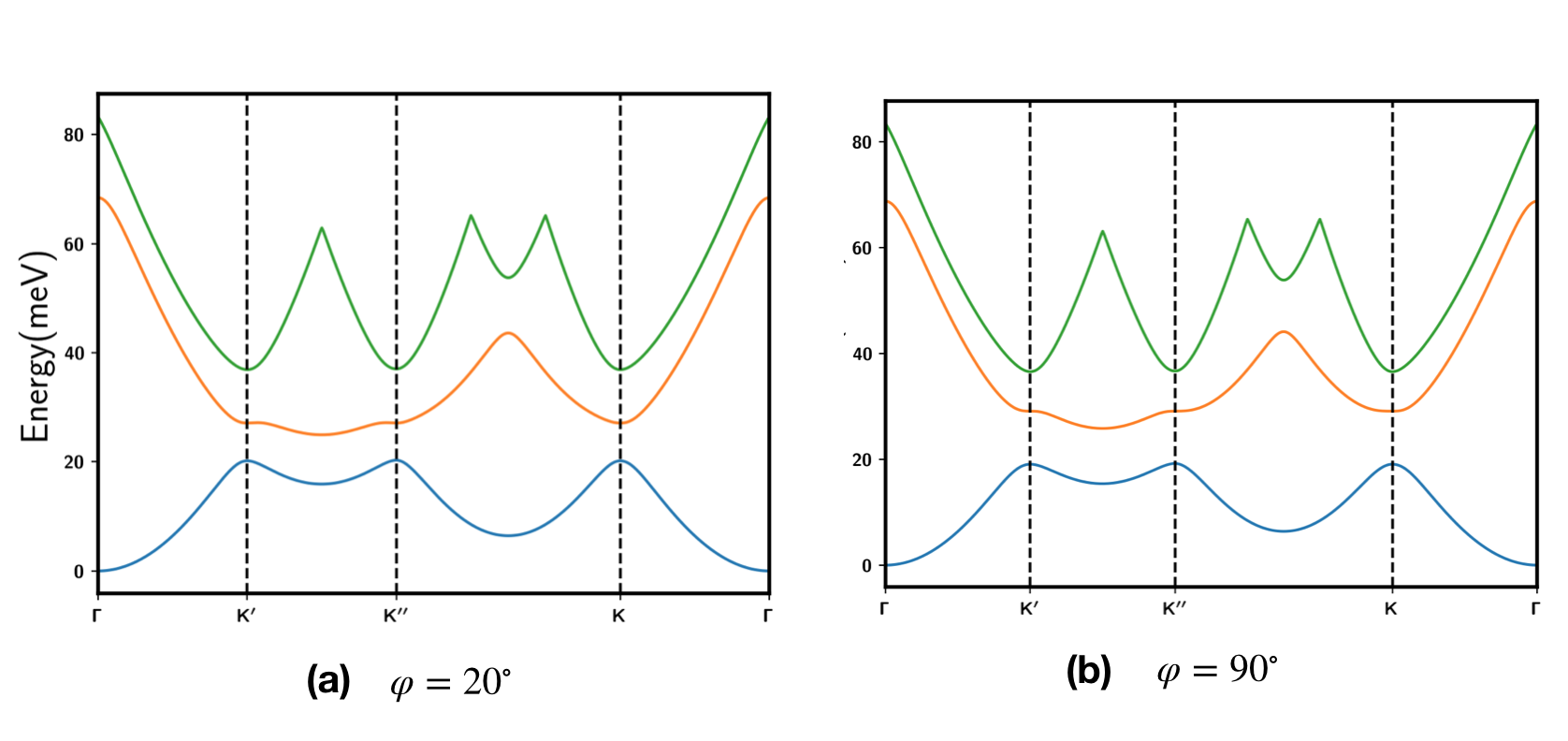}
\caption{Reconstructed moir\'e bands in the mini Brillouin zone (MBZ).  Note that we are using the hole picture here. Here we use $V=5$ meV and the Fourier components included have the form $V_{G_1} = Ve^{i\varphi}$.}
\label{fig:band}
\end{figure}

The moir\'e Hamiltonian is

\begin{equation}
  H_M= V \sum_{\mathbf k} e^{i\varphi_j} c^\dagger_{a\sigma}(\mathbf k+G_j) c_{a\sigma}(\mathbf k)
  \label{eq:moire_H}
\end{equation}
where $\mathbf G_1=(\frac{4\pi}{\sqrt{3}a_M},0)$ and $\mathbf G_2, \mathbf G_3,...,\mathbf G_6$ are generated by $C_6$ rotation of $\mathbf G_1$. Note that the $C_3$ symmetry and Hermiticity guarantees that $\varphi_j=\varphi$ for $j=1,3,5$ and $\varphi_j=-\varphi$ for $j=2,4,6$.  The time-reversal symmetry guarantees that the two valleys follow the same $H_M$. A mirror reflection symmetry relates the $H_M$ of the two layers.  In the end we only need to keep two parameters $V$ and $\varphi$ which we estimate based on the AA stacked homobilayer results \cite{wu2019topological,tang2021geometric}.   We plot the reconstructed moir\'e bands in Fig.~\ref{fig:band},  which  is not very sensitive to $\varphi$. One can see that there is a narrow band, for which we will try to build Wannier orbital. Note, there is very little Berry curvature, allowing us to obtain tightly localized Wannier orbitals.

Following the procedure\cite{zhang2019bridging} to construct Wannier orbitals and project the Coulumb interaction, we reach the following lattice model:

\begin{align}
  H&=-t \sum_{\langle ij \rangle}\sum_{a=t,b}\sum_{\sigma=\uparrow,\downarrow}c^\dagger_{i;a\sigma}c_{j;a\sigma}+\frac{U}{2}\sum_i (n_{i;t}^2+n_{i;b}^2)+U'\sum_i n_{i;t}n_{i;b}+V \sum_{\langle ij \rangle}({n_{i;t}n_{i;t}+n_{i;b}n_{j;b}})+V' \sum_{\langle ij \rangle}({n_{i;t}n_{i;b}+n_{i;b}n_{j;t}})\notag\\
  &=-t \sum_{\langle ij \rangle}\sum_{a=t,b}\sum_{\sigma=\uparrow,\downarrow}c^\dagger_{i;a\sigma}c_{j;a\sigma}+\frac{U}{2}\sum_i n_i^2+V \sum_{\langle ij \rangle}n_i n_j-\delta U\sum_i n_{i;t}n_{i;b}-\delta V \sum_{\langle ij \rangle}({n_{i;t}n_{i;b}+n_{i;b}n_{j;t}})
\end{align}

For $\theta=3.0^\circ$, we obtain the following tight binding model: $t=-2.1$ meV, $t'=0.4$ meV. By assuming the dielectric constant $\epsilon=20$ and the inter-layer distance $d=0.7$ nm, we get $U=34$ meV, $U'=26$ meV, with $\frac{\delta U}{U}=0.23$, meanwhile $V=8$ meV and $V'\approx V$ with $\frac{\delta V}{V}\approx 0.01$. Correspondingly, $\frac{\delta V}{U}=0.002$, which is quite small.   In contrast, if we take the inter-layer distance to be $d=2.5$ nm, we get $U=34$ meV, $U'=14$ meV, which gives $\frac{\delta U}{U}\approx 0.58$.  We also have $V=7$ meV with $\frac{\delta V}{V} \approx 0.2$. Thus the anisotropic terms for $d=2.5$ nm are significantly larger. 

For a Mott insulator with fixed density $n_i=\nu_T$ at each site, the anisotropic terms can be written as $H'=\delta U\sum_i P_{i;z}^2+2 \delta V \sum_{\langle ij \rangle} P_{i;z} P_{j;z}$, where $P_{i;z}=\frac{1}{2}(n_{i;t}-n_{i;b})$.

\subsection{WSe$_2$-WS$_2$-WSe$_2$ system}

Another way to realize a moir\'e bilayer is to stack two WSe$_2$ layers on top and bottom of a WS$_2$ layer in the middle. A WSe$_2$-WS$_2$ hetero-bilayer has already been experimentally demonstrated to simulate a Hubbard model and host a Mott insulator at total filling $\nu_T=1$\cite{wu2018hubbard,tang2020simulation,regan2019optical}. Basically the WS$_2$ layer provides a moir\'e superlattice potential to WSe$_2$ layer because of a small lattice constant mismatch. Meanwhile the gap of WS$_2$ is significantly larger than that of WSe$_2$, hence it also acts as an insulator for holes at the top valence band  of the WSe$_2$ layer.

The Hamiltonian of the system can be written down as

\begin{equation}
  H=H_0+H_M
\end{equation}
with
\begin{equation}
  H_0= \sum_{\mathbf k} \frac{k^2}{2m^*}c^\dagger_{a\sigma}(\mathbf k) c_{a\sigma}(\mathbf k)
\end{equation}
where $a=t,b$ labels the two layers and $\sigma=\uparrow,\downarrow$ is the spin index.

The moir\'e Hamiltonian is

\begin{equation}
  H_M= V \sum_{\mathbf k} e^{i\varphi_j}  c^\dagger_{t\sigma}(\mathbf k+G_j) c_{t\sigma}(\mathbf k)+V \sum_{\mathbf k} e^{i\varphi_j}  e^{i \mathbf{G_j}\cdot \mathbf {\delta R}}c^\dagger_{b\sigma}(\mathbf k+G_j) c_{b\sigma}(\mathbf k)
\end{equation}
where $\mathbf G_1=(\frac{4\pi}{\sqrt{3}a_M},0)$ and $\mathbf G_2, \mathbf G_3,...,\mathbf G_6$ are generated by $C_6$ rotation of $\mathbf G_1$. Note that the $C_3$ symmetry and Hermiticity guarantees that $\varphi_j=\varphi$ for $j=1,3,5$ and $\varphi_j=-\varphi$ for $j=2,4,6$.  Note the similarity of the above Hamiltonian with Eq.~\ref{eq:moire_H} of the twisted AB stacked TMD homo-bilayer, except an additional $e^{i \mathbf{G_j}\cdot \mathbf {\delta R}}$ factor due to the possible translation shift of the bottom WSe$_2$ layer relative to the top WSe$_2$ layer.  If we only consider the bottom layer, this phase factor can be gauged away by a transformation $c_{b\sigma}(\mathbf k) \rightarrow c_{b\sigma}(\mathbf k)e^{i \mathbf k \cdot \mathbf{\delta R}}$, corresponding to a translation of $\mathbf{\delta R}$.  Therefore the band structures of the two WSe$_2$ layer are the same. However, the Wannier orbitals of the two layers have a relative shift of $\mathbf{\delta R}$.  In the real experiment, $\mathbf{\delta R}$ may be random if the van der waals force between the two WSe$_2$ layers is weak due to a large layer separation. In this paper we focus on the limit that $|\mathbf{\delta R}|$ is much smaller than the moir\'e lattice constant $a_M$. In this limit the physics is very similar to the twisted AB stacked TMD homo-bilayer and the low energy physics is also captured by the anisotropic SU(4) model described in the last subsection.

 \subsection{Weak coupling limit of $\nu=\frac{3}{4}$}
\begin{figure}[ht]
\centering
\includegraphics[width=0.9\textwidth]{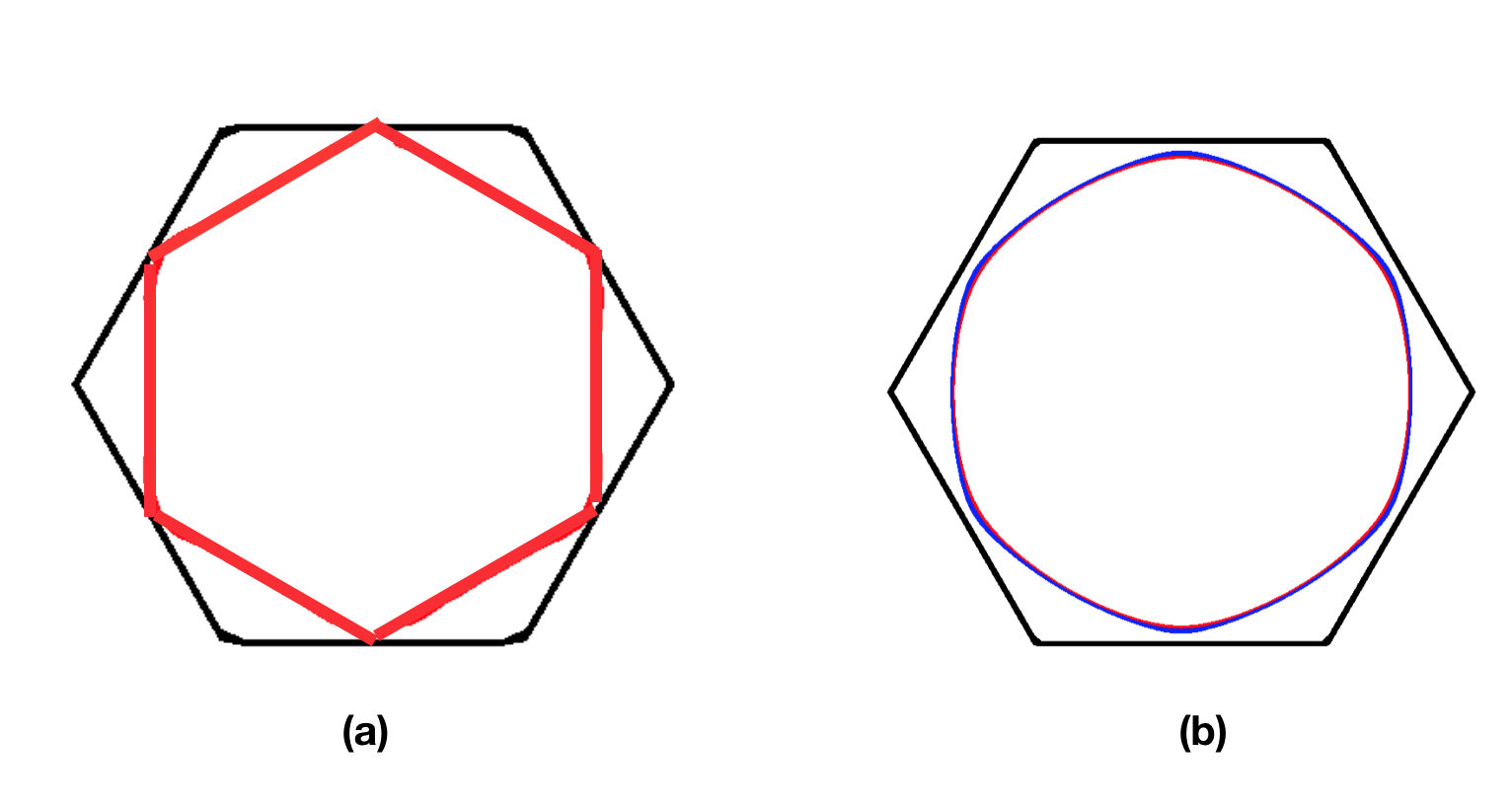}
\caption{(a) Nested Fermi surface at $U/t \rightarrow 0$ limit for the filling $\nu=\frac{3}{4}$ per flavor. There is a perfect nested condition under a shift of momentum $\mathbf M_i$, where $\mathbf M_i,\  i=1,2,3$ labels the three in-equivalent M points. (b)Fermi surface at $\nu=\frac{3}{4}$ from the calculation of the twisted AB stacked TMD with parameter $V=5$ meV, $\varphi=90^\circ$. The red and blue lines denote the two valleys respectively. There is no perfect nesting due to a small $t'$. }
\label{fig:nested_FS}
\end{figure}

 In this paper we analyzed the $U/t \gg 1$ limit at $\nu_T=3$ or $\nu=\frac{3}{4}$ per flavor of the $SU(4)$ Hubbard model. In this section we point out that even at the weak coupling limit the system is an insulator due to the perfectly nested Fermi surface. As shown in Fig.~\ref{fig:nested_FS}, the Fermi surface at free fermion level is perfectly nested under a shift of momentum $\mathbf Q=\mathbf M_i$, where $i=1,2,3$. $\mathbf M_i$ correspond to the three in-equivalent M points: $\mathbf M_1=(0,\frac{2\pi}{\sqrt{3}})$ and $\mathbf M_2, \mathbf M_3$ are generated by $C_3$ rotation. As a result, the metal phase at $U=0$ point is unstable to the formation of particle-hole order at momentum $\mathbf M_i$, which opens a charge gap.  A natural possibility is a charge-density-wave (CDW) or spin-density-wave(SDW) state with a $2\times 2$ unit cell along the $\mathbf a_1=(1,0)$ and the $\mathbf a_2=(-\frac{1}{2},\frac{\sqrt{3}}{2})$ directions.

In the spin $1/2$ Hubbard model on square lattice, the SDW phase at weak coupling just crossovers to the Neel order at the $U/t \gg 1$ limit. Our $SU(4)$ case is very different. As we showed in this paper, the $U/t \gg 1$ limit has a plaquette order, a chiral spin liquid phase and likely also a decoupled chain phase. Therefore, several phase transitions should happen when increasing $U/t$ starting from the weak coupling insulator. This offers a wonderful platform to study spin phase transitions without worrying too much about the closing of the charge gap, unlike other systems where the intermediate spin liquid phase is interrupted by the metal-insulator transition.

\section{Spin model with on-site operators}
For DMRG, we need to rewrite the ring-exchange term $P_{ij}$ and $P_{ijk}$  in terms of the on-site spin operators.  There are two different ways of representing them, which we introduce in the following.  We used both representations in our DMRG simulations. In representation I we can use three quantum numbers, while in representation II only two quantum numbers can be used.  The representation I is used in the infinite DMRG (iDMRG) code and finite DMRG code for $L_y=8$ (and some  calculations for $L_y=6$). The representation II is used by the finite DMRG code for $L_y=4,6$. In all cases finite and infinite DMRG with the two different representations give consistent results.

\subsection{Representation I}

At each site, the Hilbert space dimension is four. We label the four states as $1,2,3,4$. Then we can define $16$ operators labeled at each site $i$ as:
\begin{equation}
    S_{i;\alpha \beta}= \ket{\alpha}_i\bra{\beta}_i
\end{equation}
with $\alpha,\beta=1,2,3,4$.

In the DMRG, we use the subgroup $U(1)^3$ of $SU(4)$ to speed up the calculation. The corresponding three conserved quantum numbers are  $Q_1=\sum_i (S_{i;22}-S_{i;11})$, $Q_2=\sum_i (S_{i;33}-S_{i;11})$, $Q_3=\sum_i (S_{i;44}-S_{i;11})$.

With the above spin operators, we can rewrite the two-site ring-exchange term as
\begin{equation}
    P_{ij}= \sum_{\alpha,\beta=1,2,3,4}S_{i;\alpha \beta} \otimes S_{j;\beta \alpha}
\end{equation}

Similarly the three-site ring exchange term $P_{ijk}$ can be re-expressed as

\begin{equation}
    P_{ijk}+h.c.=\sum_{\alpha_1\beta_1;\alpha_2 \beta_2;\alpha_3 \beta_3} A_{\alpha_1\beta_1;\alpha_2\beta_2;\alpha_3\beta_3} S_{i;\alpha_1\beta_1}\otimes S_{j;\alpha_2 \beta_2} \otimes S_{k;\alpha_3\beta_3}
\end{equation}

The coefficient can be obtained in a brute force way:

\begin{equation}
    A_{\alpha_1\beta_1;\alpha_2 \beta_2;\alpha_3\beta_3}= \text{Tr}[ P_{ijk} *( S_{i;\beta_1 \alpha_1} \otimes S_{j;\beta_2 \alpha_2} \otimes S_{k;\beta_3 \alpha_3}]/\text{Tr}[(S_{i;\alpha_1 \beta_1} \otimes S_{j;\alpha_2 \beta_2} \otimes S_{k;\alpha_3 \beta_3}*( S_{i;\beta_1 \alpha_1} \otimes S_{j;\beta_2 \alpha_2} \otimes S_{k;\beta_3 \alpha_3} )]
\end{equation}
where we used the fact that $\text{Tr}[ S_{i;\alpha \beta} S_{i;\tilde \alpha \tilde \beta}]\propto \delta_{\alpha \tilde \beta}\delta_{\beta \tilde \alpha}$.

In total there are $124$ terms for $P_{ijk}+h.c.$ in the above form.  $i(P_{ijk}-h.c.)$ can be expanded with the spin operators in the same way and there are $120$ terms.

\subsection{Representation II\label{subsection:representation_II}}

 In the second representation, we view the Hilbert space of each site  as a tensor product of two independent spin $1/2$.  For example, one can view the first "spin" as the layer pseudospin and the second spin as the real spin.  We define Pauli matrix $\tau_\mu$ for the layer and Pauli matrix $\sigma$ for the spin.   Then we can define $16$ operators: $S_{\mu \nu}=\tau_{\mu}\otimes \sigma_{\nu}$ with $\mu,\nu=0,x,y,z$.

In this representation, DMRG calculation can only use two conserved numbers:  $S_{0z}, S_{z0}$.   Then we should replace $\tau_x,\sigma_x$ and $\tau_y,\sigma_y$ with $\tau_p, \sigma_p=\left(\begin{array}{cc}0 &1 \\ 0 &0 \end{array}\right)$  and $\tau_m, \sigma_m=\left(\begin{array}{cc}0 &0 \\ 1 &0 \end{array}\right)$.  We label the on-site spin operator as $S_{\alpha \beta}=\tau_{\alpha} \otimes \sigma_{\beta}$ with $\alpha,\beta=0,p,m,z$.

With the above spin operators, the two-site ring exchange term is

\begin{align}
     P_{ij}&=0.25 S_{i;00}\otimes S_{j;00}+0.5 S_{i;0p} \otimes  S_{j;0m}+0.5 S_{i;0m} \otimes S_{j;0p}+0.25 S_{i;0z}\otimes S_{j;0z}\notag\\
     &+0.5 S_{i;p0}\otimes S_{j;m0}+S_{i;pp}\otimes S_{j;mm}+S_{i;pm}\otimes S_{j;mp}+0.5 S_{i;pz}\otimes S_{j;mz}\notag\\
     &+0.5 S_{i;m0}\otimes S_{j;p0}+S_{i;mp}\otimes S_{j;pm}+S_{i;mm}\otimes S_{j;pp}+0.5 S_{i;mz}\otimes S_{j;pz}\notag\\
     &+0.25 S_{i;z0}\otimes S_{j;z0}+0.5 S_{i;zp}\otimes S_{j;zm}+0.5 S_{i;zm}\otimes S_{j;zp}+0.25 \otimes S_{i;zz}S_{j;zz}
 \end{align} 

 The first term is just a constant and in the DMRG calculation we remove it and use

 \begin{equation}
     \tilde P_{ij}=P_{ij}-0.25 S_{i;00}\otimes S_{j;00}
 \end{equation}

Similar to the previous subsection, the three-site ring exchange term can be expanded as

\begin{equation}
    P_{ijk}+h.c.=\sum_{\alpha_1\beta_1;\alpha_2 \beta_2;\alpha_3 \beta_3} A_{\alpha_1\beta_1;\alpha_2\beta_2;\alpha_3\beta_3} S_{\alpha_1\beta_1}(i) \otimes S_{\alpha_2 \beta_2}(j) \otimes S_{\alpha_3\beta_3}(k)
\end{equation}

The coefficient can be obtained in a brute force way:

\begin{equation}
    A_{\alpha_1\beta_1;\alpha_2 \beta_2;\alpha_3\beta_3}= \text{Tr}[ P_{ijk} *( S_{i;\tilde \alpha_1 \tilde \beta_1} \otimes S_{j;\tilde \alpha_2 \tilde \beta_2} \otimes S_{k;\tilde \alpha_3 \tilde \beta_3} )]/\text{Tr}[(S_{i;\alpha_1 \beta_1} \otimes S_{j;\alpha_2 \beta_2} \otimes S_{k;\alpha_3 \beta_3}*( S_{i;\tilde \alpha_1 \tilde \beta_1} \otimes S_{j;\tilde \alpha_2 \tilde \beta_2} \otimes S_{k;\tilde \alpha_3 \tilde \beta_3})]
\end{equation}
where $\tilde \alpha=0,m,p,z$ for $\alpha=0,p,m,z$.

For $P_{ijk}+h.c.$, there are $135$ non-zero terms besides the constant $S_{i;00}\otimes S_{j;00}\otimes S_{k;00}$. However, $45$ among them can be written as (and be combined with) Heisenberg coupling on one bond, i.e. it is in the form $\frac{1}{2} S_{00}(k) \otimes \tilde P_{ij}$.  So we only need to keep the remaining $90$ terms from 

\begin{equation}
    \tilde P_{ijk}+h.c.=(P_{ijk}+h.c.)-\frac{1}{2}(S_{i;00}\otimes \tilde P_{jk}+S_{j;00}\otimes \tilde P_{ki}+S_{k;00}\otimes \tilde P_{ij})-\frac{1}{8} S_{i;00}\otimes S_{j;00}\otimes S_{k;00}
\end{equation}

The chirality term $\chi_{ijk}=i(P_{ijk}-h.c.)$ can be expanded similarly and there are in total $120$ terms.

In this representation, the original Hamiltonian can be rewritten as

\begin{equation}
    H=\tilde H+\frac{3}{4} (J+K)
\end{equation}
with

\begin{equation}
    \tilde H=\tilde J \sum_{\langle ij \rangle} \tilde P_{ij} +3\tilde K \sum_{\langle ij k\rangle} (\tilde P_{ijk}+h.c.)
    \label{eq:representation_2_H}
\end{equation}
where $\tilde J=J+3K$ and $\tilde K=K$.  Thus $\frac{\tilde K}{\tilde J}=\frac{K/J}{1+3K/J}$.

In DMRG for the representation II, we simulate the Hamiltonian in Eq.~\ref{eq:representation_2_H} directly.

\subsection{Dipole-Spin representation}

In the context of moir\'e bilayer, it is useful to view the $SU(4)$ spin formed as an electric dipole moment entangled with the real spin.  More specifically, we can define layer pseudospin $\vec P_i=\frac{1}{2} \vec \tau_i$ and the real spin $\vec S_i =\frac{1}{2} \vec \sigma_i$, where $\vec \tau$ and $\vec \sigma$ are defined in the last subsection.  In this language $\vec P$ carries an electric dipole moment, while $\vec S$ carries a magnetic moment.  The spin model can be rewritten in terms of $\vec S$ and $\vec P$.

First, the Heisenberg term becomes:
\begin{equation}
  \tilde P_{ij}=\vec S_i \cdot \vec S_j +\vec P_i \cdot \vec P_j+4 (\vec S_i \cdot \vec S_j)(\vec P_i \cdot \vec P_j)
\end{equation}

The real part of the ring-exchange term is

\begin{equation}
  \tilde P_{ijk}+h.c.=-8 [\vec S_i \cdot (\vec S_j \times \vec S_k)][\vec P_i \cdot (\vec P_j \times \vec P_k)]+2 \sum_{\tilde i \tilde j \tilde k}(\vec S_{\tilde i} \cdot \vec S_{\tilde k})(\vec P_{\tilde j} \cdot \vec P_{\tilde k})
\end{equation}
where in the second term $\tilde i \tilde j \tilde k$ is summed over permutation of $ijk$.

The chirality term is

\begin{align}
  \chi_{ijk}&=i (P_{ijk}-h.c.)\notag\\
  &=\vec P_i \cdot (\vec P_j \times \vec P_k)+\vec S_i \cdot (\vec S_j \times \vec S_k)\notag\\
  &~~+4 (\vec P_i \cdot \vec P_j+\vec P_j \cdot \vec P_k+\vec P_i \cdot \vec P_k)[\vec S_i \cdot (\vec S_j \times \vec S_k)]\notag\\
  &~~+4 (\vec S_i \cdot \vec S_j+\vec S_j \cdot \vec S_k+\vec S_i \cdot \vec S_k)[\vec P_i \cdot (\vec P_j \times \vec P_k)]
\end{align}

\section{Large N mean field calculation}
We can obtain a phase diagram for SU(N) spin model at fixed filling $\nu=\frac{\nu_T}{N}=\frac{1}{4}$ by taking $N$ to infinity. We introduce fermionic spinon $f_{i;\alpha}$ at each site with $\alpha=1,2,...,N$ to denote the spin degree of freedom. The constraint is
\begin{equation}
  \sum_\alpha f^\dagger_{i;\alpha}f_{i;\alpha}=\nu_T
\end{equation}

Equivalently the density of each flavor is $\nu=\frac{\nu_T}{N}$.  For simplicity we only keep the spin model to the third order of $t/U$:
\begin{align}
H_S&=-J\sum_{\langle ij \rangle}f^\dagger_{i;\alpha}f_{j;\alpha} f^\dagger_{j;\beta} f_{i;\beta}\notag\\
&~~+3K \sum_{i,j,k\in \bigtriangleup}\left(f^\dagger_{i;\alpha}f_{k;\alpha}f^\dagger_{k;\gamma}f_{j;\gamma}f^\dagger_{j;\beta}f_{i;\beta} e^{i \Phi_3}+h.c.\right)
\label{eq:spin_model}
\end{align}
with $J=2\frac{t^2}{U}$ and $K=2\frac{t^3}{U^2}$. $\Phi_3$ is the external magnetic flux through a triangle.

We can use a mean field theory to get the phase diagram with $K$ and $\nu$. The mean field theory ignores fluctuations and is well known to fail for $SU(2)$ case. However, in the $N\rightarrow \infty$ limit the fluctuation around the saddle point is suppressed by the integration of fermions and mean field theory becomes more accurate.  

To have a controlled large $N$ calculation, we need to rescale $\tilde K=KN^2$ and $\tilde J=JN$. We will fix the filling $\nu=\nu_T/N=\frac{1}{4}$, $\tilde K$ and $\tilde J$ when taking $N$ to infinity.

We can decouple the spin model to have a mean field ansatz:
\begin{equation}
  H_M=-\sum_{\langle ij \rangle}\chi_{ij} f^\dagger_{i\alpha}f_{j\alpha}+h.c.-\sum_i \mu_i f^\dagger_{i;\alpha}f_{i;\alpha}
  \label{eq:mean_field}
\end{equation}
where $\mu_i$ is introduced to satisfy the constraint: $\frac{1}{N}\sum_\alpha \langle f^\dagger_{i;\alpha}f_{i;\alpha}\rangle=\nu$.

Mean field ansatz $\chi_{ij}$ can be obtained by Feynman's variational principle\cite{brinckmann2001renormalized}. Basically the free energy $\beta F\leq \Phi[\chi_{ij}]$, here
\begin{equation}
\Phi=\langle S -\tilde S \rangle- \log \tilde Z 
\end{equation}
where $S$ is the action of the full Hamiltonian and $\tilde S$ is the action corresponding to the mean field ansatz in Eq.~\ref{eq:mean_field}. $\tilde Z=\int D[f] e^{-\beta \tilde S}$ is the partition function of the mean field theory.  

We can obtain mean field ansatz $\chi_{ij}$ by minimizing $\Phi[\chi_{ij}]$, which leads to self consistent equations:

\begin{equation}
  \chi_{ij}=\tilde J\langle T_{ji} \rangle+3\tilde K \sum_{k,i,j \in \bigtriangleup}e^{-i \Phi_3}\langle T_{jk} \rangle\langle T_{ki} \rangle 
  \label{eq:self_consistent_equations}
\end{equation}
where $T_{ij}=\frac{1}{N}\sum_\alpha f^\dagger_{i\alpha} f_{j\alpha}$.

At $T=0$, variational energy is:
\begin{equation}
  \frac{E_M}{N}=-\tilde J\sum_{\langle ij \rangle}\langle \hat T_{ij} \rangle \langle T_{ji} \rangle-3\tilde K\sum_{i,j,k\in \bigtriangleup}(e^{-i\Phi_3} \langle T_{ik}\rangle \langle T_{kj} \rangle  \langle T_{ji}\rangle +h.c.)
  \label{eq:mean_field_energy}
\end{equation}

In the calculation, we choose different unit cell with size $m\times n$ and solve the self-consistent equations in Eq.~\ref{eq:self_consistent_equations} using the iteration method starting from a randomly chosen ansatz.  The iteration method is not guaranteed to find the global minimum. We need to start from several different initial ansatz and keep the best solution.   During every step of the iteration, $m\times n$ number of chemical potentials are solved to implement the constraint that $\frac{1}{N}\langle f^\dagger_{i;\alpha} f_{i;\alpha} \rangle=\nu$  at each site $i$.

From numerical simulation, we find three phases by varying $\frac{\tilde K}{\tilde J}$: (I) A plaquette order when $\frac{\tilde K}{\tilde J}<0.233$; (II) A SU(4)$_1$ chiral spin liquid when $\frac{\tilde K}{\tilde J}\in (0.233,0.767)$; (III) A decoupled chain phase when $\frac{\tilde K}{\tilde J}>0.767$.  Using $\frac{K}{J}=\frac{1}{N}\frac{\tilde K}{\tilde J}=\frac{1}{4} \frac{\tilde K}{\tilde J}$, we obtain the phase diagram in terms of $K/J$ for the SU(4) model as shown in the main text.

\begin{figure}[ht]
\centering
\includegraphics[width=0.95\textwidth]{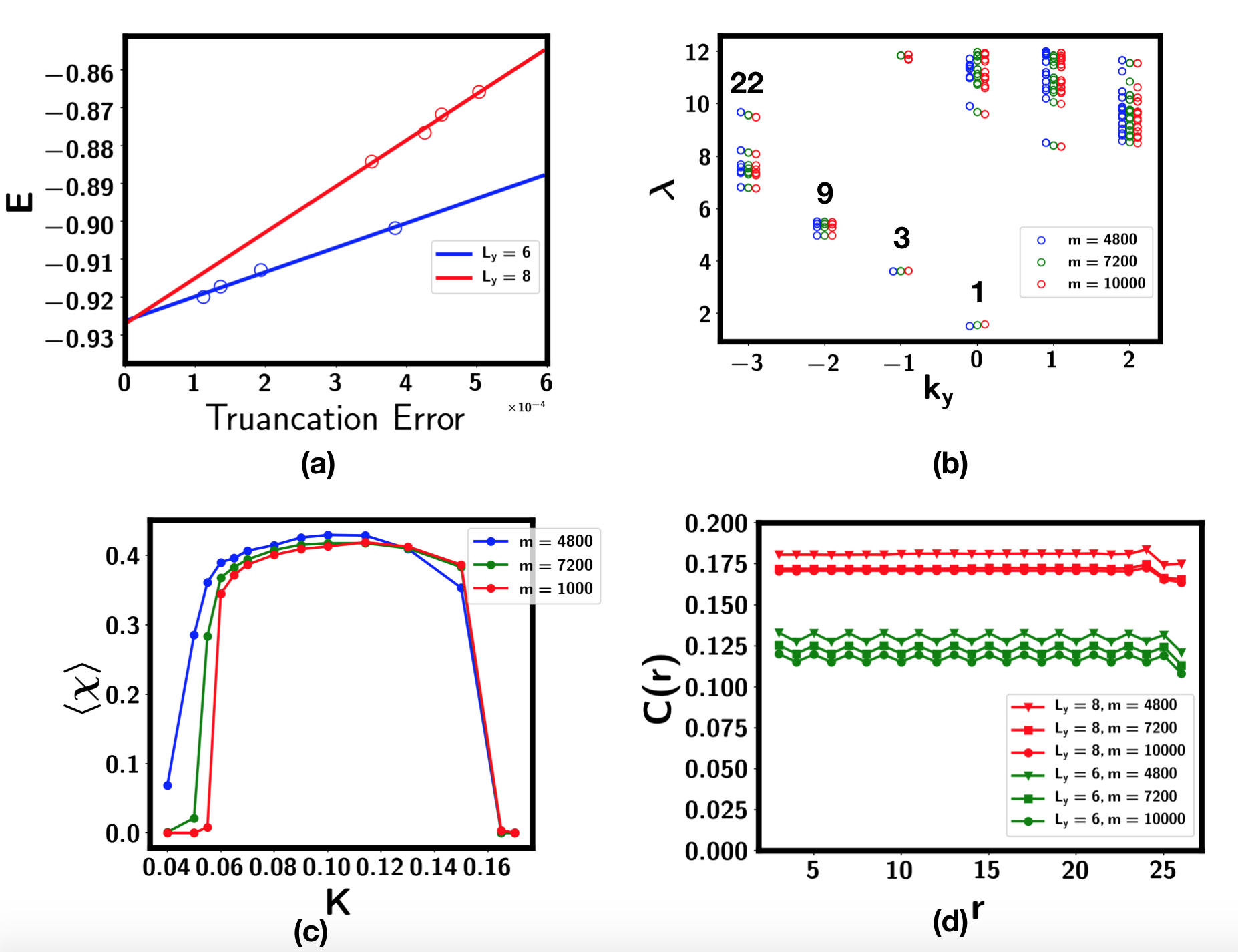}
\caption{Finite DMRG results with varying bond dimension $m$. (a) Energy vs truncation error for $L_y=6$ and $L_y=8$ at $K/J=0.114$. For $L_y=6$, the bond dimension used is $m=2400,4800,7200,10000$. For $L_y=8$, the bond dimension used is $m=4800,6000,7200,10000$. (b) The entanglement spectrum for bond dimension $m=4800,7200,10000$ for $L_y=6$ at $K/J=0.114, \Phi=\frac{\pi}{16}$. (c)Chirality order vs $K/J$ for $L_y=8$ at various bond dimensions, obtained using complex code. (d) Chiral-chiral correlation function with various bond dimensions at $L_y=6,8$ at $K/J=0.114$.  }
\label{fig:convergence}
\end{figure}

\section{Convergence of DMRG}

In this section we show more data to demonstrate the convergence of the finite DMRG when increasing the bond dimension.  As shown in Fig.~\ref{fig:convergence}(a), the energy extrapolated to zero truncation error limit is quite close for $L_y=6$ and $L_y=8$ systems, although the $L_y=8$ has larger truncation error with the same bond dimension.  The entanglement spectrum at $L_y=6$ basically does not change when changing the bond dimension from 4800 to 10000.  In Fig.~\ref{fig:convergence}(c) we show the chirality order vs K for $L_y=8$ at various bond dimensions. One can see that a small bond dimension  overestimates the chirality order strength at the lower critical point $K\approx 0.055$, while it tends to underestimate the chirality order strength at the higher critical point $K\approx 0.165$.  In Fig.~\ref{fig:convergence} we show that the chiral-chiral correlation functions converges quite well with bond dimension $10000$ for both $L_y=6$ and $L_y=8$.  One interesting feature is that the chiral order correlation function has an oscillation with $2\times 1$ unit cell for $L_y=6$, which is stable when increasing the bond dimension. Such an oscillation is absent for $L_y=4,8$. One explanation of the oscillation is the following: when $L_y$ is not a multiple of $4$, a translation invariant state can not be gapped due to the LSM constraint. Hence the system may want to enlarge the unit cell to make the number of fundamental representations within the $2\times L_y$ unit cell as a multiple of $4$. There is no such issue for $L_y=4 n$.  We believe the oscillation is an artifact for finite $L_y \notin 4 Z$  and  will be  absent at the $L_y\rightarrow \infty$ limit.

\begin{figure}[ht]
\centering
\includegraphics[width=0.95\textwidth]{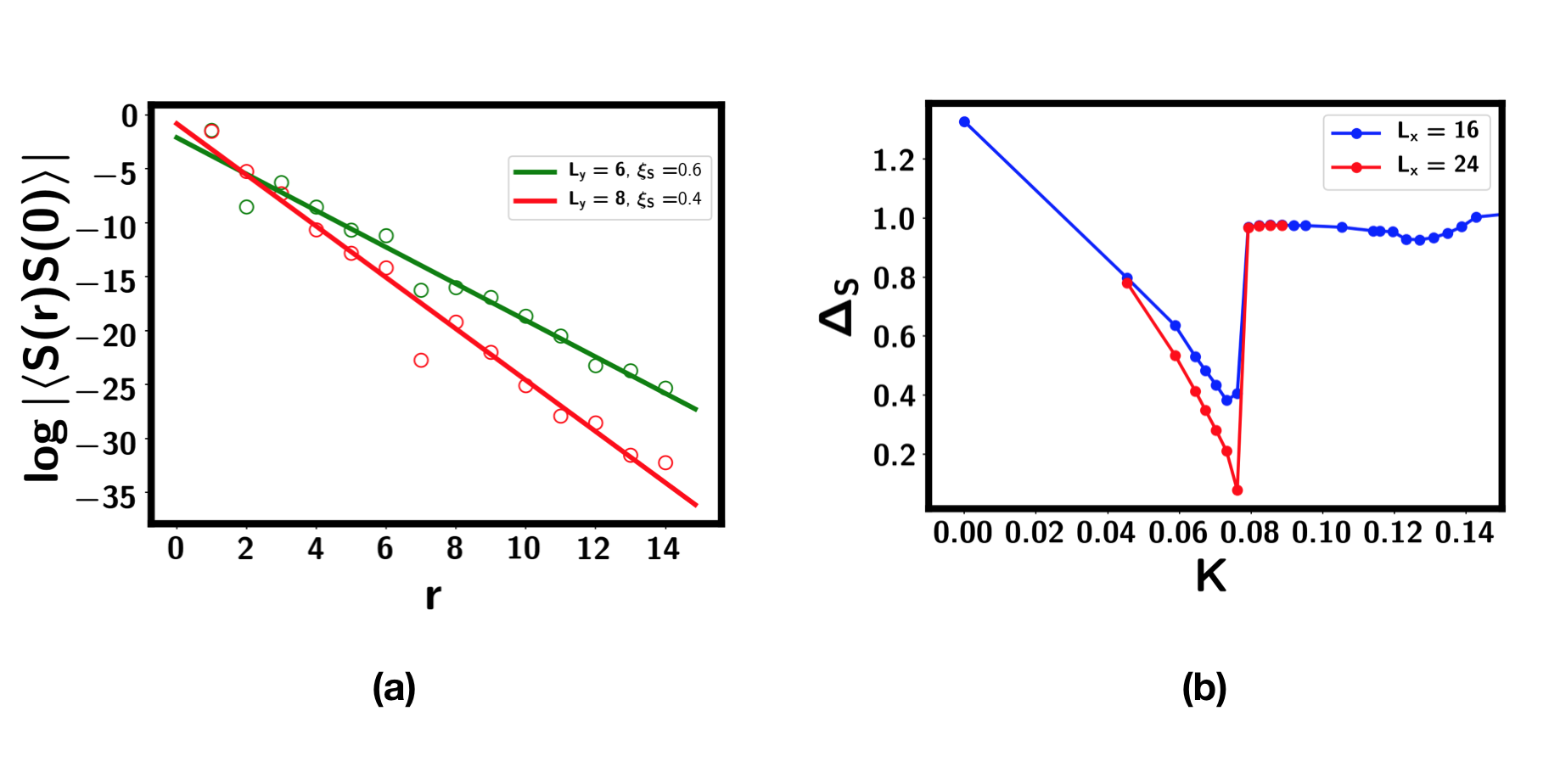}
\caption{(a) Spin-spin correlation length for $L_y=6,8$ at bond dimension $m=10000$ for $K=0.114$. One can see that the correlation length is smaller than one lattice space. (b) Spin gap $\Delta_S$ for $L_y=4$. The spin gap is defined as the lowest energy in the sector of the adjoint representation of SU(4) relative to the ground state in the SU(4) singlet sector. In DMRG using only Abelian symmetry, we just need to focus on the sector with a spin flip. In the dipole-spin representation with two spin $1/2$ $\vec P$ and $\vec S$, the excitation is in the sector $P_z=1$ or $S_z=1$.}
\label{fig:spin_correlation}
\end{figure}

In addition to the bond dimension, another controlling parameter in our calculation is $L_y$. The largest $L_y$ we can reach is 8 and one may wonder whether it is large enough to obtain the property of the system at the 2D limit.  In Fig.~\ref{fig:spin_correlation}(a) we fit the correlation length in the chiral spin liquid phase from the spin-spin correlation function. The correlation length is $0.6$ and $0.4$ for $L_y=6$ and $L_y=8$ respectively, which is consistent with a large spin gap $\Delta_S \sim J$ as shown in Fig.~\ref{fig:spin_correlation}(b).  $L_y=6$ and $L_y=8$ are already much larger than the correlation length and therefore we believe the phase at $L_y=6,8$ should already be very close to the 2D limit.

\section{Plaquette order at the Heisenberg limit}

In this section we  discuss the Heisenberg limit at $K=0$. There is an apparent difference between the even $L_y$ and the odd $L_y$. For odd $L_y$, even if we have a $2\times 1$ enlarged unit cell, the system is still  gapless as required by the Lieb-Schultz-Mattis (LSM) theorem. Therefore, strictly speaking, odd $L_y$ is not very informative.  In the following we will discuss the even and odd $L_y$ cases separately. For even $L_y$, we always find a gapped phase with strong translation symmetry breaking.  For odd $L_y$, there is also translation symmetry breaking, though it is weaker than the even $L_y$ case.  Meanwhile, the system is gapless, but the central charge does not scale with $L_y$.  So even for the odd $L_y$ case, it is not in a uniform spin liquid phase as suggested by Ref.~\onlinecite{keselman2020emergent}.  Putting them together, our numerical results are more consistent with a picture with gapped  crystallized phase, which has additional gapless modes for odd $L_y$.

\begin{figure}[ht]
\centering
\includegraphics[width=0.6\textwidth]{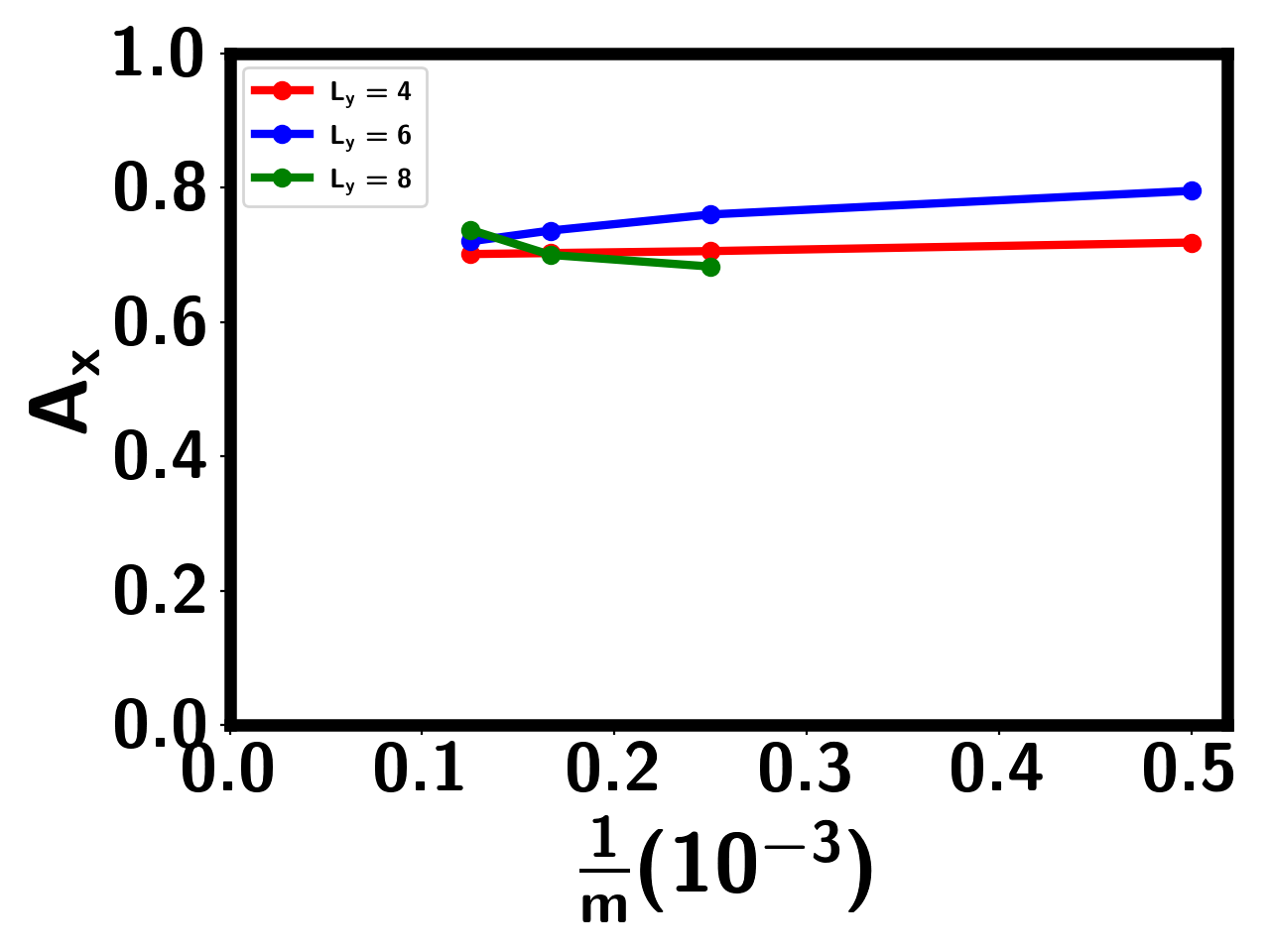}
\caption{Anisotropy $A_x=\frac{ \sum_{i_x \in \text{even}}\big(\langle \tilde P_{i,i+\hat{x}}\rangle-\langle \tilde P_{i+\hat{x},i+2\hat{x}}\rangle\big)}{\sum_{i_x \in \text{even}}\langle \tilde P_{i,i+\hat{x}}\rangle}$ at $K=0$. Here $\tilde P_{ij}=P_{ij}-\frac{1}{4}I$. $m$ is the bond dimension. One can see that the dimerization along $x$ direction remains strong when we increase $L_y$ from $4$ to $8$. The bond dimension $m$ is varied from $2000$ to $8000$.}
\label{fig:anisotropy}
\end{figure}

\subsection{Even $L_y$}

\begin{figure}[ht]
\centering
\includegraphics[width=0.95\textwidth]{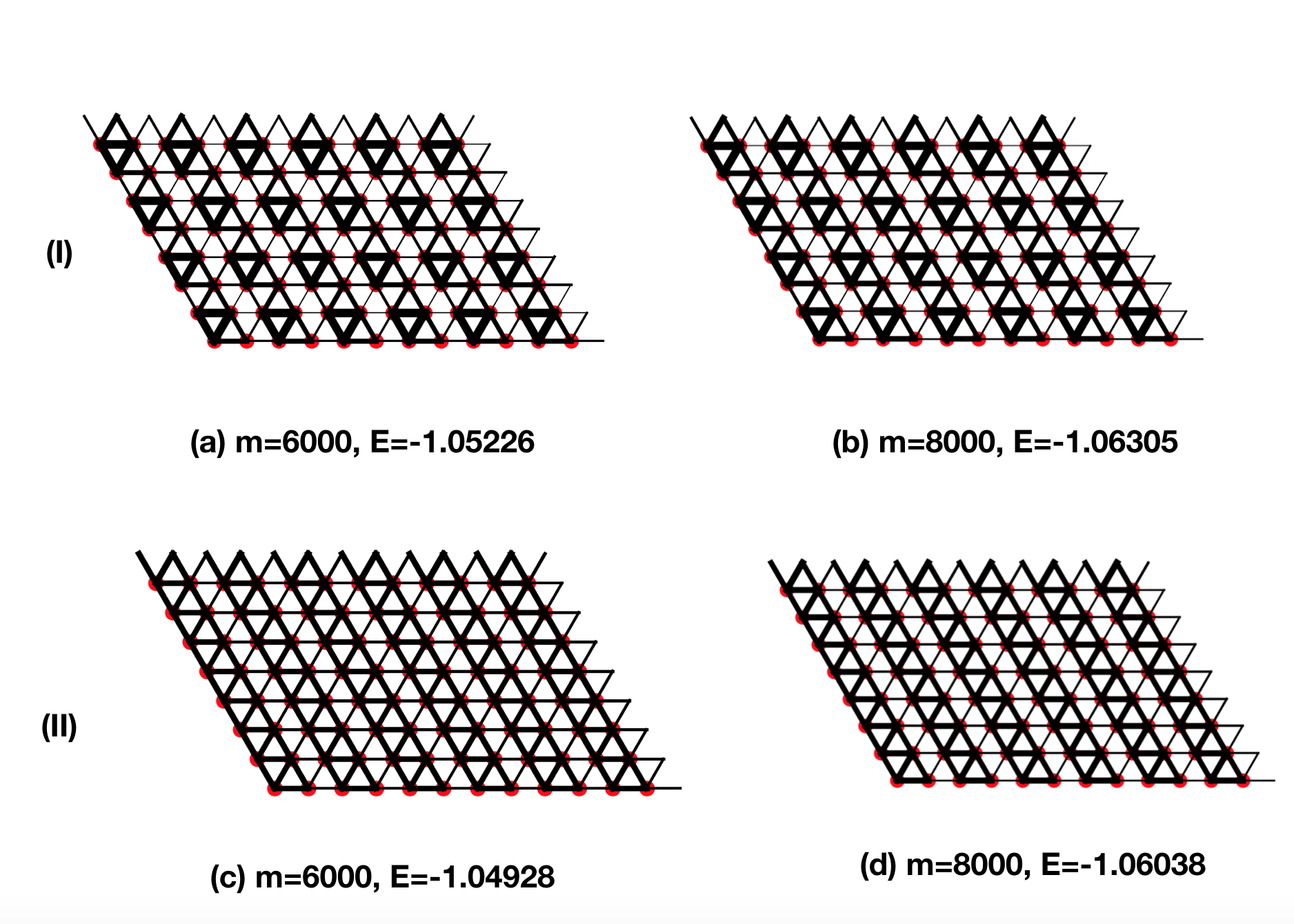}
\caption{Bond order $|\langle \tilde P_{ij}\rangle|$ for $L_y=8$ obtained by iDMRG. The first and second row correspond to two different states obtained from different initial ansatz. $m$ is the bond dimension and $E$ is the energy per site in unit of $J$.}
\label{fig:L8_bond}
\end{figure}

 First, as shown in Fig.~\ref{fig:anisotropy}, there is a strong dimerization along the $x$ direction for $L_y=4,6,8$. For $L_y=4,6$, the unit cell is $2\times 1$. For $L_y=8$, there is also a translation symmetry breaking along the $y$ direction, resulting in a $2\times 2$ unit cell.  Spin crystal phase with $2\times 2$ unit cell is found as ground state in the large N mean field calculation\cite{zhang_2020,yao2020topological}, while the crystal phase with $2\times 1$ unit cell as found in our DMRG calculation for $L_y=4,6$ is shown to be a competing state\cite{yao2020topological}. Therefore our DMRG result agrees with the large $N$ mean field analysis.  The same dimerization pattern was reported for $L_y=4$ in DMRG study of  Ref.~\onlinecite{keselman2020emergent}, but it was interpreted as a backscattering instability of a translation invariant spinon Fermi surface state\cite{keselman2020emergent}. We note that the spinon Fermi surface state has a weak instability only at $L_y=2$ and the instability should decay quickly when increasing $L_y$. In contrast, the dimerization along $x$ direction in our DMRG remains strong when we increase $L_y$ from $4$ to $8$, which is inconsistent with the spinon fermi surface picture. Therefore, we interpret our numerical discovery as a strong translation symmetry breaking order, along the same spirit of the crystal phase in the large $N$ calculation.

\begin{figure}[ht]
\centering
\includegraphics[width=0.6\textwidth]{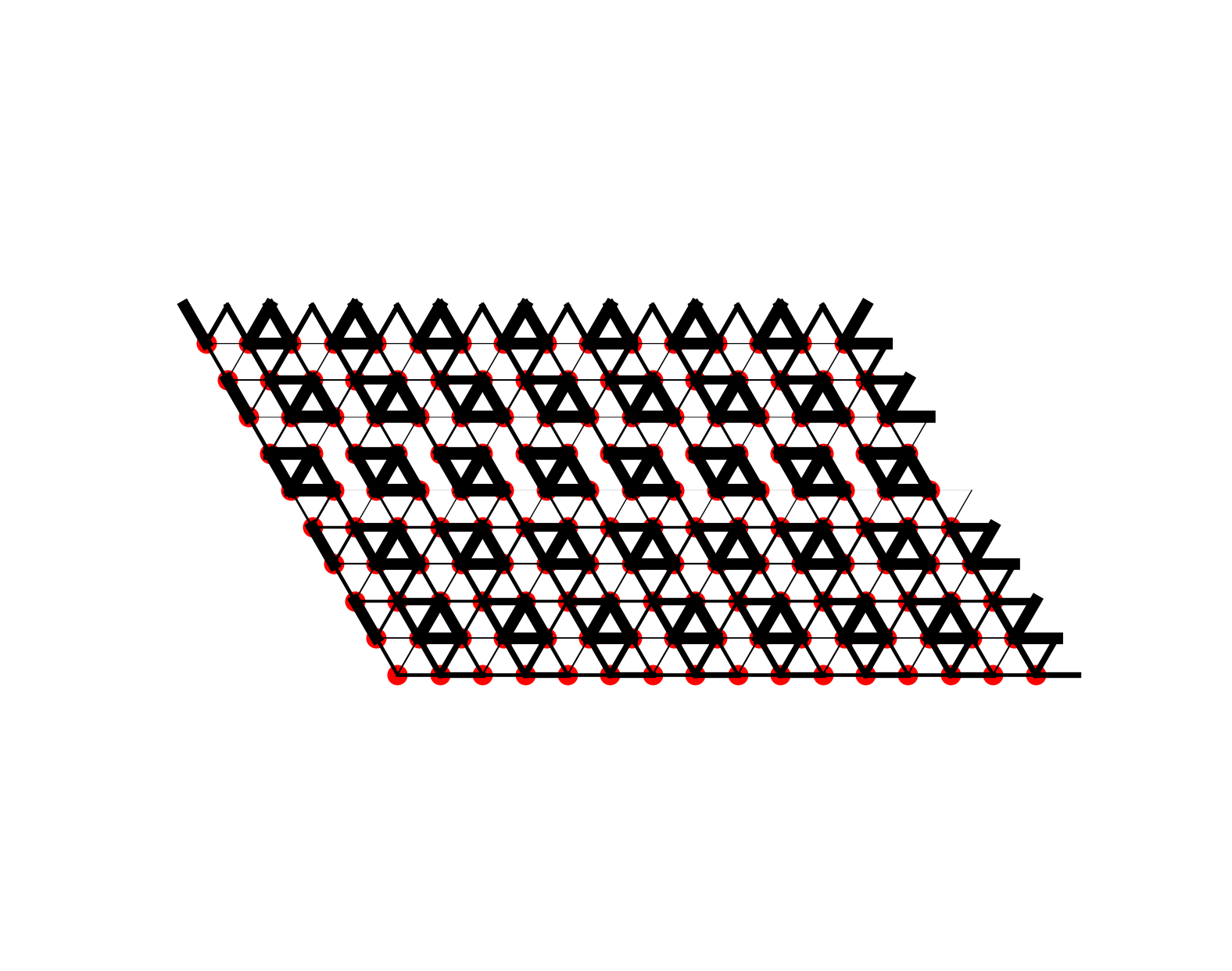}
\caption{Bond order $|\langle \tilde P_{ij}\rangle|$ for $L_y=10$ for bond dimension $m=7000$ with iDMRG. The energy is $E=-0.98543 J$. One can see clear plaquette order, though there is domain in $y$ direction. We tried different randomized initial ansatz, but the results are similar. We failed to obtain a perfect plaquette order without domain.}
\label{fig:L10_bond}
\end{figure}

As the dimerization in $x$ direction is strong for  $L_y=4,6,8$, we can start from a picture of decoupled two-leg stripe along $y$ direction, whose length is equal to $L_y$. The two-leg stripe can remain translation invariant along $y$ direction when $L_y$ is small, but it is known to be unstable to plaquette order when $L_y \rightarrow \infty$\cite{keselman2020emergent}. Indeed, we find weak dimerization along $y$ direction for $L_y=8$ and $L_y=10$. First, in Fig.~\ref{fig:L8_bond} we show the bond order for two different states obtained from two different randomized initial ansatz at $L_y=8$.  One can see that a state with $2\times 2$ unit cell has slightly smaller energy than the $2\times 1$ stripe state.  We note that for $L_y\leq 6$, different initial states always lead to the same $2\times 1$ stripe phase. Our results suggest that the crystal phase with $2\times 2$ unit cell has higher energy than the $2\times 1$ stripe  when $L_y<8$ and becomes competitive at $L_y=8$.  We expect it to become more favored at larger $L_y$. To test this, we show the bond order for $L_y=10$ obtained with bond dimension $m=7000$ in Fig.~\ref{fig:L10_bond}. We can see that there is a clear plaquette ordering, though there is domain in $y$ direction, presumably because the bond dimension is still too small for $L_y=10$ and the DRMG is stuck in a local minimum. Note that in the large $N$ mean field calculation, state with any plaquette covering is degenerate to each other. The energy cost of domain wall is obtained in $1/N$ expansion and may still be small for $N=4$. Despite that the DMRG may not be well converged to a global minimum, the result is consistent with the formation of plaquette order.
 
\begin{figure}[ht]
\begin{subfigure}{.8\textwidth}
  \centering
  % include first image
  \includegraphics[width=.8\linewidth]{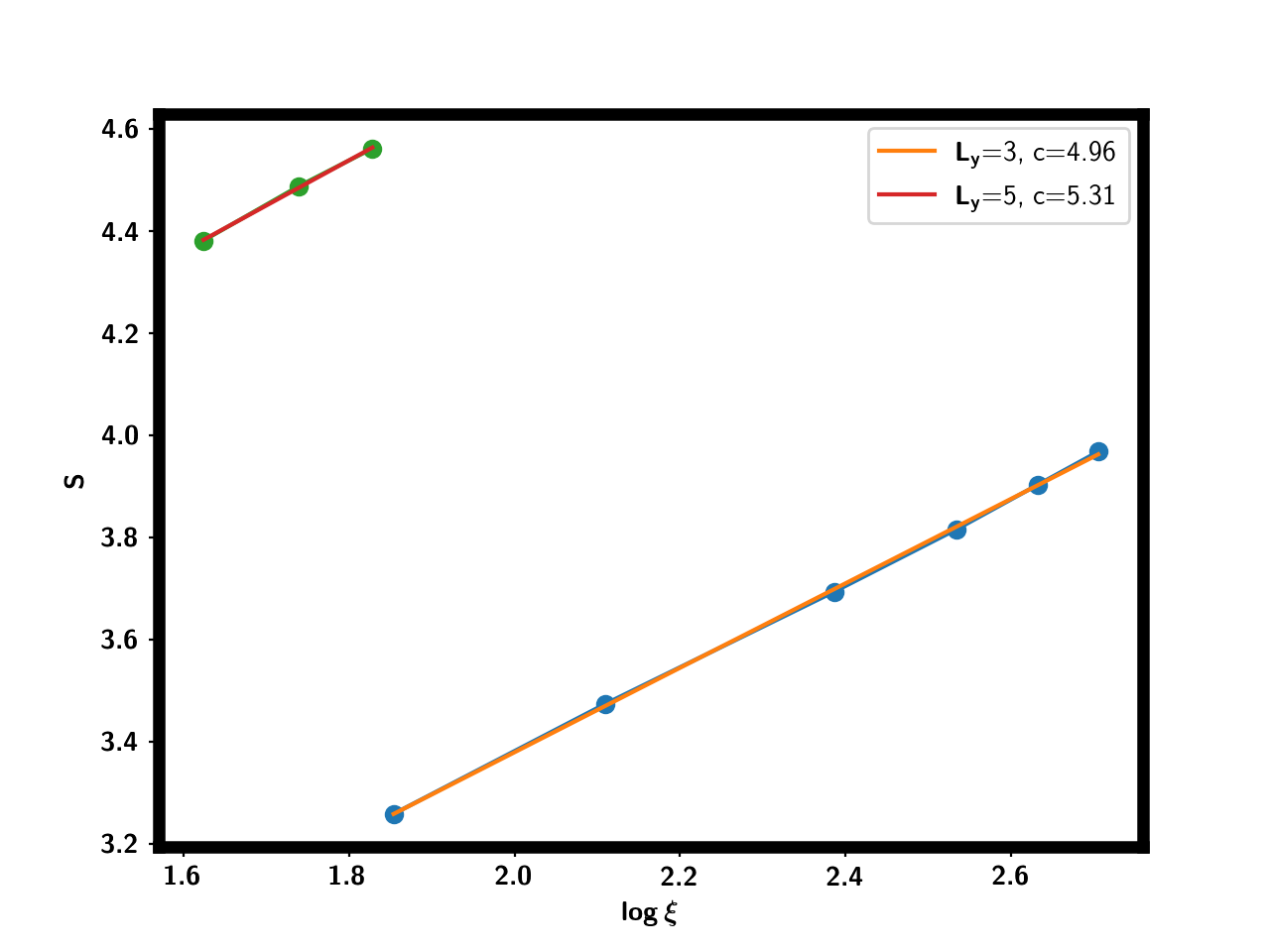}
  %\caption{Central charge fit from $S=\frac{c}{6} \log \xi$ for $L_y=3,5$.}
\end{subfigure},

\begin{subfigure}{.49\textwidth}
  \centering
  % include second image
  \includegraphics[width=.8\linewidth]{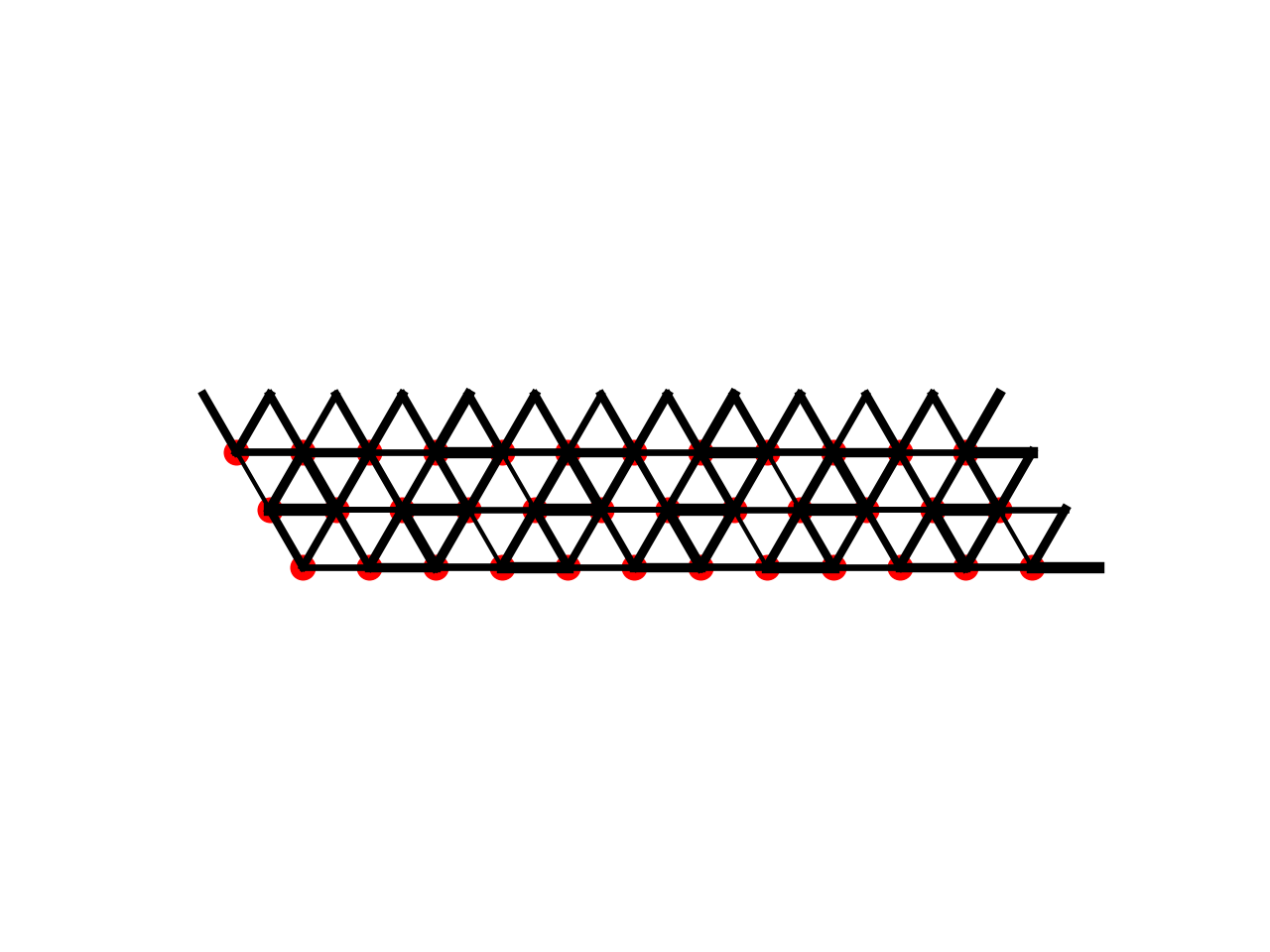}
  %\caption{Bond order $\langle \tilde P_{ij}\rangle $ for $L_y=3.}
\end{subfigure}
\begin{subfigure}{.49\textwidth}
  \centering
  % include second image
  \includegraphics[width=.8\linewidth]{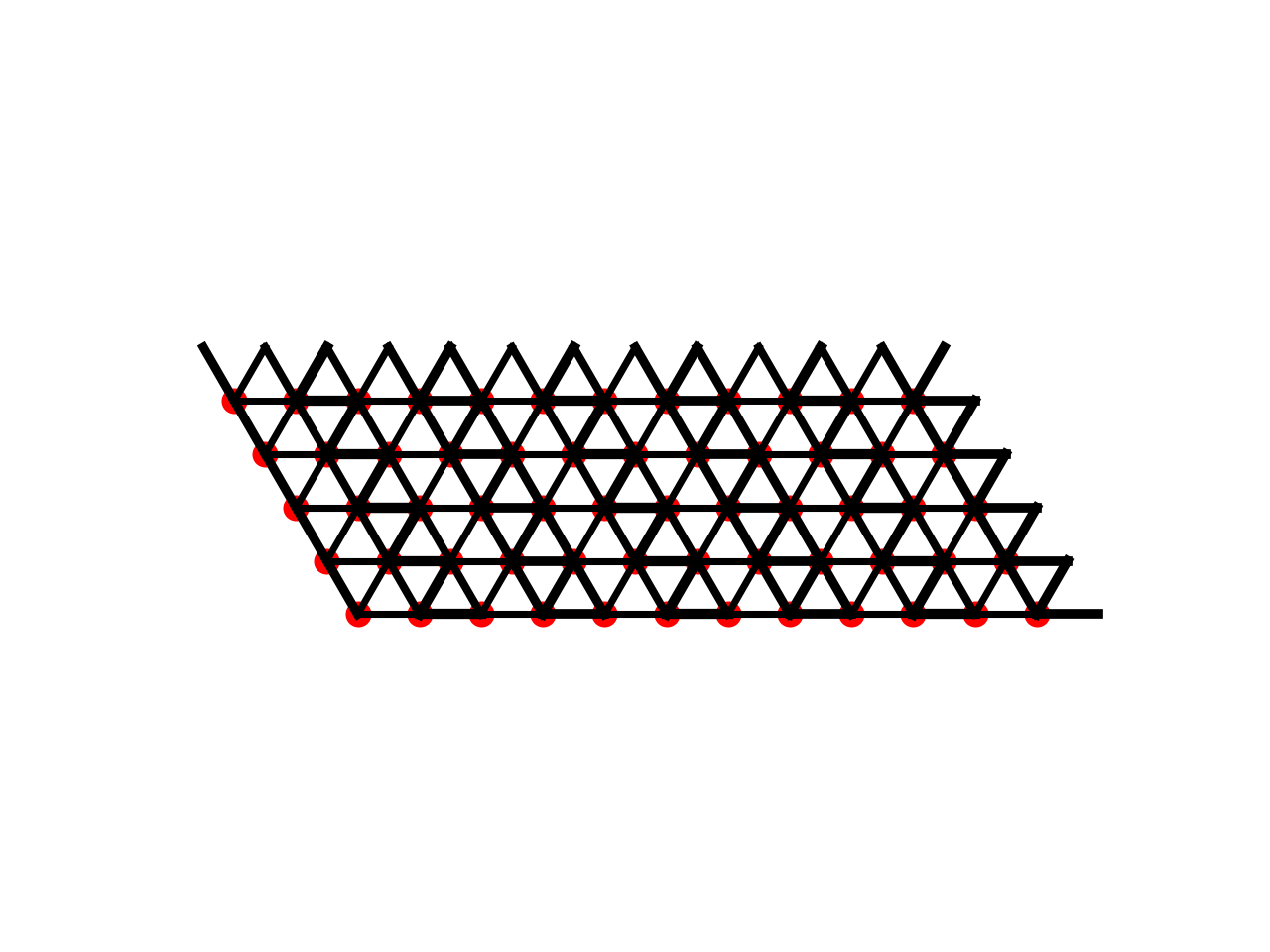}
  %\caption{Bond order $\langle \tilde P_{ij}\rangle $ for $L_y=5.}
\end{subfigure}
\caption{(a) Central charge fit from $S=\frac{c}{6} \log \xi$ for $L_y=3,5$; (b)Bond order $\langle \tilde P_{ij} \rangle$ for $L_y=3$;  (c) Bond order for $L_y=5$.}
\label{fig:odd_Ly}
\end{figure}

\subsection{Odd $L_y$}

In Fig.~\ref{fig:odd_Ly} we show our results for $L_y=3,5$. First, for both cases we find translation symmetry breaking into a $2\times 1$ unit cell. Especially, the $L_y=5$ case has the same pattern as those found in $L_y=4,6$.  Meanwhile, the system is gapless with a central charge close to $c\approx 5$ for both $L_y=3,5$.  Even for $L_y=3$, this central charge is smaller than the $SU(8)_1$ theory suggested by Ref.~\onlinecite{keselman2020emergent}, which should give $c=7$. Therefore we conclude that even  $L_y=3$  is not in a simple uniform spin liquid phase. For $L_y=5$, the central charge does not become larger, in contradiction with a spinon Fermi surface state, which requires the central charge to increase linearly with $L_y$.  In summary, our results align with the picture of a gapped crystallized phase.  The gapless mode with finite central charge is required by the LSM theorem for odd $L_y$ even with a $2\times 1$ enlarged unit cell. Such gapless mode does not exist in even $L_y$ cases and is probably just an unnecessary complexity with odd $L_y$.

\section{Variational wavefunction of the CSL}

The $SU(4)_1$ CSL reported in our DMRG calculation was actually also found in a large N mean field calculation\cite{zhang_2020}.   The mean field theory is based on the standard Abrikosov fermionic parton representation of the spin operator:
\begin{equation}
  S_{i;\alpha \beta}=f^\dagger_{i;\alpha}f_{i;\beta}
\end{equation}
where $\alpha,\beta=1,2,3,4$ is the $SU(4)$ flavor index.

The mean field ansatz of the parton $f$ can be obtained self consistently. In the intermediate regime of $K/J$, one finds the following ansatz:
\begin{equation}
  H_M=- t_f \sum_{\langle ij \rangle}e^{i \varphi_{ij}} f^\dagger_{i;\alpha}f_{j;\alpha}-\mu \sum_i f^\dagger_{i;\alpha}f_{i;\alpha}
\end{equation}
Here $\varphi_{ij}$ is the phase associated with the hopping of the spinon $f$. The amplitude of the hopping is found to be uniform.  The flux of the hopping is on average $\frac{2\pi}{4}$ per unit cell. However, we find that non-uniform flux state is also energetically favorable. We have a $2\times 2$ enlarged unit cell. Let us assume the flux in these four unit cells as $\Phi_1,\Phi_2,\Phi_3,\Phi_4$. Although $\Phi_1+\Phi_2+\Phi_3+\Phi_4=2\pi$, each of them is not $\frac{2\pi}{4}$ and the exact value depends on the value of $K/J$. This suggests that the resulting CSL phase may have a $2\times 2$ unit cell in terms of the chirality order.

With this ansatz, there are four separated Chern bands. At filling $\nu_T=1$, we can occupy the lowest band completely and each spinon $f_{\alpha}$ is in a Chern insulator state with $C=1$. We can write down a variational wavefunction as a generalization of the traditional Gutzwiller projection method:

 \begin{equation}
     \ket{\Psi}=P_{G} \prod_{\alpha=1,2,3,4} \text{Slater}[f_\alpha] 
 \end{equation}
 where $P_{G}$ is the Gutzwiller projection to fix the constraint:
 
 \begin{equation}
     \sum_{\alpha=1,2,3,4}f^\dagger_{i;\alpha} f_{i;\alpha}=1
 \end{equation}
 
 Each $\text{Slater}[f_\alpha]$ is a slater determinant for the fermion $f_\alpha$ following the mean field ansatz. The above wavefunction can be simulated by the standard variational Monte Carlo technique.  $\varphi_{ij}$ can be viewed as variational parameters to be determined by VMC.

\section{Derivation of the topological field theory for the CSL}
\label{Sec:SupD}

In this section we try to derive a Chern-Simons theory for the $SU(4)_1$ CSL. As described in the last section, in the mean field level, we have each fermionic spinon $f_{\alpha}$ in a $C=1$ Chern insulator. Next we couple $f_{\alpha}$ to an $U(1)$ gauge field $a$ coming from the constraint $\sum_{\alpha}f^\dagger_{i;\alpha}f_{i;\alpha}=1$. We get the action

\begin{align}
    L=L[f_1,a+A_{s1}+A_{s2}+A_{s3}]+L[f_2,a+A_{s1}-A_{s2}-A_{s3}]+L[f_3,a-A_{s1}+A_{s2}-A_{s3}]+L[f_4,a-A_{s1}-A_{S2}+A_{s3}]
\end{align}
Here $A_{s1},A_{s2},A_{s3}$ is the external spin field corresponding to $S_{z0},S_{0z},S_{zz}$ respectively.

Each $L[f_I,a]$ is describing an integer quantum Hall effect coupled to gauge field $a$,  it is:
\begin{equation}
    L[f_I,a]=-\frac{1}{4\pi}\alpha_I d \alpha_I +\frac{1}{4\pi} a d \alpha_i
\end{equation}

Putting them together, we get:

\begin{align}
    L&=-\frac{1}{4\pi} \sum_{I=1}^4 \alpha_I d \alpha_I + \frac{1}{2\pi} 
    \sum_{I=1}^4 a d \alpha_I+\frac{1}{2\pi} (A_{s1}+A_{s2}+A_{s3}) d \alpha_1 +\frac{1}{2\pi} (A_{s1}-A_{s2}-A_{s3}) d \alpha_2 \notag\\
    &+\frac{1}{2\pi} (-A_{s1}+A_{s2}-A_{s3}) d \alpha_3+\frac{1}{2\pi} (-A_{s1}-A_{s2}+A_{s3}) d \alpha_4
\end{align}
 Next we integrate $a$, which locks $\alpha_1+\alpha_2+\alpha_3+\alpha_4=0$.  We can then substitute $\alpha_4=-(\alpha_1+\alpha_2+\alpha_3)$ and get
 
 \begin{align}
     L=-\frac{2}{4\pi} \sum_{I=1}^3 \alpha_I d \alpha_i -\frac{1}{4\pi}\sum_{I\neq J} \alpha_I d \alpha_J+\frac{1}{2\pi}A_{s1} d (2\alpha_1+2\alpha_2)+\frac{1}{2\pi}A_{s2} d (2\alpha_1+2\alpha_3)+\frac{1}{2\pi}A_{s3} d (-2\alpha_2-2\alpha_3)
 \end{align}
 
 We can rewrite it as
 
 \begin{equation}
     L=-\frac{1}{4\pi} \sum_{IJ}\alpha_I K_{IJ} d\alpha_J+\frac{1}{2\pi} \sum_{iI}A_{s_i} q_{iI} d \alpha_I
 \end{equation}
 where $\alpha=(\alpha_1, \alpha_2, \alpha_3)$ is a three dimension vector, thus $I,\,J\in \{1,\,2,\,3\}$ and wee have the 3x3 $K$ matrix:
 \begin{equation}
     K=\begin{pmatrix} 2 & 1 &1 \\ 1 &2 &1 \\ 1 &1 &2 \end{pmatrix}
     \label{eq:K_matrix}
 \end{equation}
 
 And the charge  $q_{iI}$ matrix is:
  \begin{equation}
     q=\begin{pmatrix} 2 & 2 & 0 \\ 2 &0 &2 \\ 0 & -2 & -2 \end{pmatrix}
     \label{eq:q_matrix}
 \end{equation}
 equivalently the charge vectors for $A_s =\left (A_{s1}, A_{s2}, A_{s3} \right )$ are $q_1=(2,2,0)$, $q_2=(2,0,2)$, $q_3=(0,-2,-2)$, 
 Then Hall conductivity is
 \begin{equation}
     \sigma_{ij}= q_{iI} \left [ K^{-1} \right ]_{IJ} q^T_{Jj}=4 \delta_{ij}
 \end{equation}

 where summation over repeated indices is assumed. The CSL contains four anyons, with statistics 
 \begin{equation}
 \theta=0,\, \frac{3}{4}\pi, \,\pi, \, \frac{3}{4} \pi
 \label{Eq:theta}
 \end{equation}. 
 
{Note, if we redefine $\tilde{K} = SKS^T$, where the similarity transform $$S = \begin{pmatrix} 1 & 0 & 0 \\ 0 & -1 & 0 \\ 0 & 1 & -1 \end{pmatrix} $$ has $Det[S]=1$, 
 the resulting $\tilde{K}$ matrix is precisely the Cartan matrix\cite{Georgi:1982jb} of SU(4), i.e.  $ \tilde{K}=\begin{pmatrix} 2 & -1 &0 \\ -1 &2 &-1 \\ 0 & -1 &2 \end{pmatrix}$ revealing the underlying connection to SU(4)$_1$ topological order.}
 
 Let us now discuss how the full SU(4) global symmetry acts on the anyons. The action of the U(1) subgroups are captured by the charge matrix $q$. Briefly, the two anyons with topological spin $\theta_a = \theta_{\bar{a}}= \frac{3\pi}{4}$ transform as (anti) fundamental representations i.e. the $\bf 4$ and $\bf \bar{4}$ representations. On the other hand the third anyon (fermion) is a bilinear of the $a,\,\bar{a}$ anyons, hence transform as the $\bf 6$ or $\bf 10$ representation of SU(4). This also counts as fractionalization, since there are no local excitations that are also electrically neutral, transforming in these representations.
 
 Finally, let us mention a simple way to view this topological order and symmetry action in terms of the Kitaev 16 fold way, in particular the $\nu=6$
member. Recall, the 16-fold way is a sequence of topological orders that partly mimic a Z$_2$ toric code topological order, in containing a fermion, but host different numbers of chiral Majorana edge modes $\nu$ and thus edge chiral central charge $c=\nu/2$. Note, that at $\nu=6$ this corresponds exactly to the edge central charge of our theory. For even integer $\nu$, the two nontrivial quasiparticles besides the fermion have topological spin $c\frac{\pi}{4}$. For $c=3$ this gives us exactly our anyon content.  The SU(4) symmetry can also be implemented in an elegant way - using the relation between the Lie algebras $SU(4)~ SO(6)$, we note that the six chiral Majorana modes at the edge transform as the vector representation of $SO(6)$, while the Z$_2$ fluxes for the fermions, bind six Majorana zero modes and hence transform as the spinor representations {\bf 4} or $\bf {\bar{4}}$. These Z$_2$ fluxes correspond to the anyons $a,\,\bar{a}$. Their topological spin is readily computed from noting that it must be the same as a $\pi$ flux in an integer quantum Hall state with $\sigma_{xy} = c$. The effective action $L = \frac1{4\pi}\sum_{i=1}^ca_ida_i +\frac1{2\pi}Ada_i$. Setting $dA/2\pi = \frac12$ this implies a `charge' vector $1/2$ for each component and a $K_Q=\mathcal I$ identity K-matrix , leading to a topological spin: $\theta = \pi\sum_{i=1}^c\frac12\cdot \frac12=\frac{c\pi}4$. Finally, setting $c=3$ matches the statistics of the anyons in eqn \ref{Eq:theta}.

In Table.~\ref{table:csl} we list some key differences between this $SU(4)_1$ CSL and the familiar $SU(2)_1$ CSL in the spin $1/2$ systems.

\begin{table}
\centering
\begin{tabular}{ |c||c|c|} 
 \hline
 Property & SU(2)$_1$  CSL  & SU(4)$_1$ CSL  \\ 
 \hline
 Number of quasi-particles & 2 & 4 \\ 
 Edge Central Charge (c) & 1 & 3 \\
  ES Degeneracy & 1, 1, 2, 3, 5 $\dots$ & 1, 3, 9, 22 $\dots$ \\
 \hline
\end{tabular}
\caption{Comparison between the $SU(4)_1$ CSL and the familiar $SU(2)_1$ CSL in spin $1/2$ systems.}
 \label{table:csl}
 \end{table}

\section{Spin flux insertion}

\subsection{A simple version: $S_{z0}$ insertion}

We can use one of the three conserved charges to do a flux insertion and detect spin Hall conductivity. Let us use $U(\varphi)=e^{i \frac{1}{2} S_{z0} \varphi}$. So we will impose the boundary condition that $S(\mathbf r+L_y \mathbf a_2))= U^\dagger(\varphi)  S(\mathbf r) U(\varphi)$. The spin operators change as

\begin{align}
   & S_{0\mu} \rightarrow  S_{0\mu} \notag\\
    & S_{z\mu} \rightarrow  S_{z\mu} \notag\\
    & S_{p\mu} \rightarrow  e^{-i  \varphi}S_{p\mu} \notag\\
     & S_{m\mu} \rightarrow  e^{i \varphi}S_{m\mu} \notag\\
\end{align}
where $\mu=0,p,m,z$.

In the Hamiltonian, we will have terms $S_i S_j S_k$. Whenever one of them cross the boundary, we should replace it with the above transformation.  In this way we generate a new Hamiltonian $H(\varphi)$.  $H(\varphi=2\pi)=H(\varphi=0)$. For CSL, the state $\ket{\varphi=2\pi} \neq \ket{\varphi=0}$.  We can calculate the conserved charge on one half of the system as $Q(\varphi)$, where $Q=S_{z0}, S_{0z}, S_{zz}$.  $S_{z0}$ should increase by $2$ when $\varphi$ increases to $2\pi$.  $S_{0z},S_{zz}$ should remain unchanged.

\subsection{Combined flux insertion}

We try to determine all of $K$ matrix entries through spin pumping. The $SU(4)_1$ CSL is described by a $3\times 3$ K matrix as derived in the last section. To simplify the charge vector, we can do a redefinition of the probing gauge fields:

 \begin{align}
     \tilde A_1 &= 2(A_{s1}+A_{s2}) \notag\\
     \tilde A_2 &= 2(A_{s1}-A_{s3}) \notag\\
     \tilde A_3 &= 2(A_{s2}-A_{s3}) 
 \end{align}

This is equivalent to use a new definition of the conserved charges:

 \begin{align}
     \tilde Q_1&=\frac{1}{4}(S_{z0}+S_{0z}+S_{zz}) \notag\\
     \tilde Q_2&=\frac{1}{4}(S_{z0}-S_{0z}-S_{zz}) \notag\\
     \tilde Q_3&=\frac{1}{4}(-S_{z0}+S_{0z}-S_{zz}) 
 \end{align}

 With the new basis, $\tilde A_I$ is the gauge field corresponding to the charge $\tilde Q_I$.  We can rewrite the action as
 
 \begin{equation}
     L=-\frac{1}{4\pi} \alpha^T K \alpha+\frac{1}{2\pi} \sum_{I=1}^3 \tilde A_I d \alpha_I
 \end{equation}
 
 Basically $\tilde A_1, \tilde A_2, \tilde A_3$ now can be viewed as elementary gauge field, similar to three gauge field of a three-layer quantum Hall system. The above action is actually the same theory to describe a three component bosonic quantum Hall state.
 
 If we measure $\sigma^{xy}$ in terms of these $\tilde A$, we should get a $3*3$ matrix corresponding to $\sigma^{xy}_{IJ}$.  This matrix should be exactly the same as $K^{-1}$. To measure $\sigma^{xy}_{IJ}$, we need to pump $U_J$ corresponding to $\tilde A_J$ and measure the charge $\tilde Q_I$.

 The three independent pumpings are:
 \begin{align}
 U_1(\varphi)&=e^{i \frac{1}{4} (S_{z0}+S_{0z}+S_{zz}) \varphi} \notag\\
 U_2(\varphi)&=e^{i \frac{1}{4} (S_{z0}-S_{0z}-S_{zz}) \varphi} \notag\\
 U_3(\varphi)&=e^{i \frac{1}{4} (-S_{z0}+S_{0z}-S_{zz}) \varphi}
 \end{align}
 
 One can prove that $U_I(\varphi=2\pi)=e^{-i\frac{2\pi}{4}} I$ in the four dimensional Hilbert space.  This is exactly the $Z_4$ flux shared by $SU(4)$ and the global $U(1)$.

 For each pumping $I=1,2,3$, we make a transformation:
 
 \begin{equation}
     S(\mathbf r+L_y \mathbf{a_2})=U^\dagger_I(\varphi) S(\mathbf r) U_I(\varphi)
 \end{equation}

We can get the Hall conductivity as
 
 \begin{equation}
     \tilde \sigma^{xy}_{ij}=\tilde Q_i
 \end{equation}
 where $Q_i$ is the pumped charge for pumping $U_j(\varphi=2\pi)$.

 In this process we can get a $3\times 3$ matrix $\tilde \sigma^{xy}$, which should be equal to $K^{-1}=\begin{pmatrix} \frac{3}{4} & -\frac{1}{4} & -\frac{1}{4}\\ -\frac{1}{4} & \frac{3}{4} & -\frac{1}{4} \\ -\frac{1}{4} & -\frac{1}{4} & \frac{3}{4} \end{pmatrix}$. This is confirmed in Fig.3(b) of the main text.

\section{Entanglement spectrum of the $SU(4)_1$ CSL}

Another way to characterize the CSL is in terms of its chiral edge mode through the bulk-boundary correspondence. The edge mode is described by the  $SU(4)_1$ conformal field theory. In this section we derive the spectrum of the edge theory, which should show up in the entanglement spectrum.

The CFT has a chiral central charge $c=3$, described by three independent bosons. In terms of $\varphi=(\varphi_1, \varphi_2, \varphi_3)^T$, we have the action of the edge theory:

\begin{equation}
    L=  \partial_t \varphi^T  K \partial_x \varphi
\end{equation}
where $K$ is defined in Eq.~\ref{eq:K_matrix}.

 The $3\times 3$ $K$ matrix We can rewrite it as $K=M M^T$ with

\begin{equation}
    M=\begin{pmatrix}-\sqrt{\frac{4}{3}} & \sqrt{\frac{1}{2}} & -\sqrt{\frac{1}{6}} 
    \\-\sqrt{\frac{4}{3}} & -\sqrt{\frac{1}{2}} & -\sqrt{\frac{1}{6}} \\
    -\sqrt{\frac{4}{3}} & 0 & \sqrt{\frac{2}{3}} 
    \end{pmatrix}
\end{equation}

and 

\begin{equation}
    (M^T)^{-1}=\begin{pmatrix}
    -\frac{1}{\sqrt{12}}& \frac{1}{\sqrt{2}} & -\frac{1}{\sqrt{6}}\\
    -\frac{1}{\sqrt{12}}& -\frac{1}{\sqrt{2}} & -\frac{1}{\sqrt{6}}\\
    -\frac{1}{\sqrt{12}}& 0 & \frac{2}{\sqrt{6}}
    \end{pmatrix}
\end{equation}

Then we can rewrite the theory as
\begin{equation}
    L=  \partial_t \tilde \varphi^T \partial_x \tilde \varphi
\end{equation}
with
\begin{equation}
    \tilde \varphi=M^T \varphi
\end{equation}
Noe that $e^{i l^T \varphi}= e^{i l^T (M^T)^{-1} \tilde \varphi}$ generates anyons.  It is easy to find the mutual statistics through OPE:

\begin{equation}
    e^{i l^T \varphi}  e^{i \varphi^T l'}=e^{i l^T (M^T)^{-1} \tilde \varphi}(z) e^{i \tilde \varphi^T M^{-1} l'}(w) \sim  (z-w)^{l^T (M M^T)^{-1} l'}e^{i l^T (M^T)^{-1} \tilde \varphi+i \tilde \varphi^T M^{-1} l'}(w)
\end{equation}
from which we can read out the mutual statistics to be $\theta_{l l'}=2 \pi l^T K^{-1} l'$. Note here that the simple OPE relation only applies to $\tilde \varphi$ instead of $\varphi$ because the action is the simple diagonal form only for $\tilde \varphi$.

We have the primary field $V_l=e^{i l^T \varphi}$, it can be rewritten as $V_l=e^{i l^T (M^T)^{-1}\tilde \varphi}$. Let us label $R_1=\sqrt{12}$, $R_2=\sqrt{2}$ and $R_3=\sqrt{6}$, then
\begin{equation}
    V_l=e^{i\big(\frac{p_1}{R_1}\tilde \varphi_1+\frac{p_2}{R_2} \tilde \varphi_2+\frac{p_3}{R_3} \tilde \varphi_3\big)}
\end{equation}
with
\begin{equation}
    p_1=-(l_1+l_2+l_3)\ \ p_2=l_1-l_2\ \ \ p_3=2l_3-l_1-l_2
\end{equation}

From $q^T K^{-1} l$ we know the quantum numbers are:

\begin{align}
S_{z0}&=l_1+l_2-l_3\notag\\
S_{0z}&=l_1-l_2+l_3 \notag\\
S_{zz}&=l_1-l_2-l_3
\end{align}

Finally, the scaling dimension of these primary fields and their descendants are:

\begin{equation}
    L_0=\frac{p_1^2}{2 R_1^2}+\frac{p_2^2}{2R_2^2}+\frac{p_3^2}{2R_3^2}+\sum_{i=1,2,3}\mu_i=\frac{s_1^2+s_2^2+s_3^2}{8}+\sum_{i=1,2,3}\mu_i
    \label{eq:L_0}
\end{equation}

where $\mu_i=\sum_{k\in Z^+} k n_i(k)$ is from the usual mode expansion.  $s_1=S_{z0}$, $s_2=S_{0z}$ and $s_3=S_{zz}$ label the three quantum numbers.

Note, in the entanglement spectrum, we find various Schmidt eigenvalues for a given momentum $k_y$ around the cylinder. The degeneracy of these states is obtained from the degeneracy of $L_0$, i.e. the number of state with a fixed scaling dimension.  $L_0$ has two terms: the first one is uniquely determined by the quantum number $(s_1,\,s_2,\,s_3)$.  The second one is from the usual mode expansion. If we fix the quantum number $(s_1,\,s_2,\,s_3)$, then the degeneracy is purely determined by the second term $\sum_{i=1,\,2,\,3}\mu_i$.   For each $i=1,2,3$, it is easy to find that $\mu_i=0,1,2,3,4,5,...$ has degeneracy $1,1,2,3,5,7,...$.   From this, we can find that the degeneracy for $L_0=0,1,2,3,...$ is $1,3,9,22,...$ for each fixed $(s_1,s_2,s_3)$.

For each spin sector with quantum numbers to be $(s_1,\,s_2,\,s_3)$, the degeneracy of the spectrum is always in the sequence $1,3,9,22,...$, but the starting energy (or equivalently the momentum) depends on the quantum numbers: $L_0=\frac{s_1^2+s_2^2+s_3^2}{8}$.   For $L_0=0$, the spin sector can only be $(0,\,0,\,0)$.  For $L_0=1$, all of possible states form the adjoint representation of $SU(4)$ with dimension $15$. In terms of the quantum numbers $(s_1,s_2,s_3)$, these $15$ states are: $3(0,0,0)$ and 12 additional states:$(\pm 2,\, \pm 2,\, 0),\,(\pm 2,\, 0,\,\pm 2),\,(0,\,\pm 2,\,\pm 2)$.

%,(2,0,-2),(2,-2,0),(0,2,-2),(-2,2,0),(-2,0,2),\\(0,-2,2),(0,-2,-2),(-2,0,-2),(-2,-2,0)$.

\section{Effect of the anisotropy terms}

For the integer filling $n=1$, at the $U>>t$ limit, we restrict to the Hilbert space with $n_i=1$. This is possible only for $n_{i;t}=1,n_{i;b}=0$ or $n_{i;t}=0,n_{i;b}=1$. In any case we have $n_{i;t}n_{i;b}=0$ and thus the $\delta U$ term vanishes after projecting to the restricted Hilbert space. The $\delta U$ term can come into the spin model only through $t/U$ expansion. It can modify $J\sim \frac{t^2}{U}\rightarrow \frac{t^2}{U \pm \delta U}=J \pm \delta J$, where $\frac{\delta J}{J} \sim \frac{\delta U}{U}$.  

\begin{figure}[ht]
\centering
\includegraphics[width=0.8\textwidth]{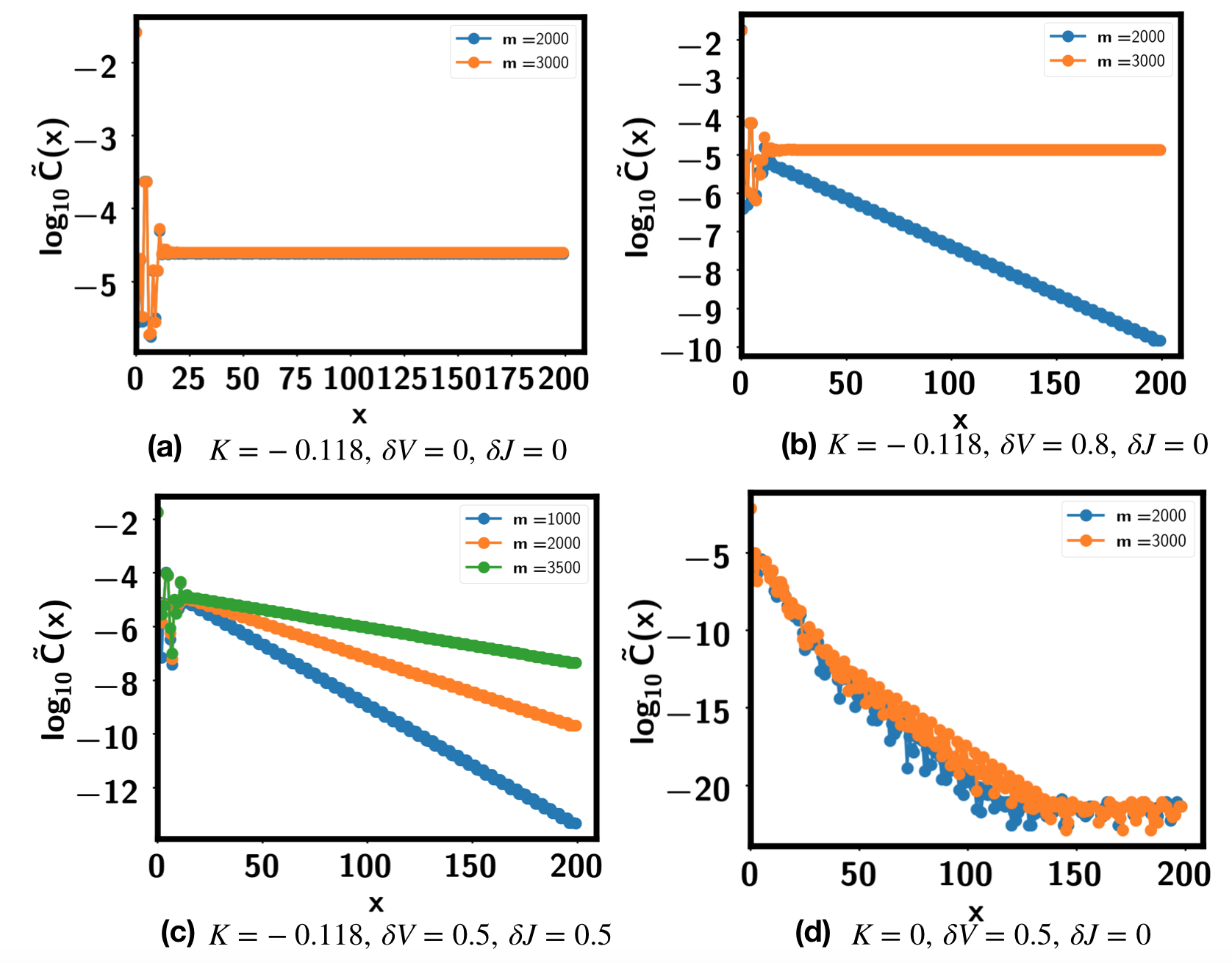}
\caption{The correlation function of the chirality order at various values of $\delta V,\delta J$ at $K=-0.118$. $m$ is the bond dimension of the iDMRG. We find that the CSL is stable at least up to $\delta V=0.8$ when $\delta J=0$. With both $\delta J=0.5,\delta V=0.5$, we believe the ground state is still a CSL. Note here the chirality-chirality correlation becomes  {\em longer ranged} when we increase the bond dimension $m$, implying truly long range order at the $m\rightarrow \infty$ limit.  Here $\tilde C(r)=\langle \tilde \chi(x) \tilde \chi(0)\rangle$ and $\tilde \chi_{ijk}=i (S_{i;11} \otimes (S_{j;12} \otimes S_{k;21}- S_{j;21}\otimes S_{k;12})$. To reduce the time cost of the computation, we only includes $2$ terms among the $120$ terms in the full expression of the chirality order $\chi_{ijk}=i(P_{ijk}-h.c.)$. As a result, $\tilde C(r)$ is order $10^{-4}$ smaller than the full correlation function $C(r)=\langle \chi(r) \chi(0) \rangle$.  A plot for $K=0,\delta V=0.5,\delta J=0$ is also shown  to indicate that there is no chiral order at $K=0$. }
\label{fig:dV}
\end{figure}

Let us derive the anisotropic term carefully. For simplicity we restrict to the second order perturbation, It is easy to derive that:

\begin{align}
  &H_S=J\sum_{\langle ij \rangle} (P_{ij}-I)\notag\\
  &+\delta J\sum_{\langle ij \rangle}[\big(S_{i;13}\otimes S_{j;31}+S_{i;14}\otimes S_{j;41}+S_{i;23}\otimes S_{j;32}+S_{i;24}\otimes S_{j;42}+S_{i;31}\otimes S_{j;13}+S_{i;32}\otimes S_{j;23}+S_{i;41}\otimes S_{j;14}+S_{i;42}\otimes S_{j;24}\big)\notag\\
  &~~-\big(S_{i;11}\otimes S_{i;33}+S_{i;11}\otimes S_{j;44}+S_{i;22}\otimes S_{j;33}+S_{i;22}\otimes S_{j;44}+S_{i;33}\otimes S_{j;11}+S_{i;33}\otimes S_{j;22}+S_{i;44}\otimes S_{j;11}+S_{i;44}\otimes S_{j;22}\big)]
\end{align}
where $J=2 \frac{t^2}{U}$ and $\delta J=2 \frac{t^2}{U'}-2\frac{t^2}{U} \approx \frac{\delta U}{U}J$ when $\delta U<<1$.

In contrast, the $\delta V$ term does not vanish after the projection. After including the $\delta V$ term, we get:

\begin{align}
  &H_S=J\sum_{\langle ij \rangle} (P_{ij}-I)\notag\\
  &+\delta J\sum_{\langle ij \rangle}\big(S_{i;13}\otimes S_{j;31}+S_{i;14}\otimes S_{j;41}+S_{i;23}\otimes S_{j;32}+S_{i;24}\otimes S_{j;42}+S_{i;31}\otimes S_{j;13}+S_{i;32}\otimes S_{j;23}+S_{i;41}\otimes S_{j;14}+S_{i;42}\otimes S_{j;24}\big)\notag\\
  &-(\delta J+\delta V) (n_{i;t}n_{j;b}+n_{i;b}n_{j;t})
\end{align}

We can also rewrite it in terms of the dipole spin $\vec P$ and the real spin $S$:

\begin{equation}
  H_S=\delta J \sum_{\langle ij \rangle} (P_{i;x}P_{j;x}+P_{i;y}P_{j;y})(4\vec{S}_i\cdot \vec{S}_j+S_{i;0}S_{j;0})+2(\delta J+\delta V)\sum_{\langle ij \rangle}P_{i;z}P_{j;z}
\end{equation}

For AB stacked TMD homo-bilayer at twist angle $\theta =3.0^\circ$, we estimate that $U/t \approx 17$ (with $\epsilon=20$). Then $J\approx 2 \frac{t^2}{U}\approx 0.007 U \approx 0.23$ meV. We have anisotropic terms   $\delta J \approx 0.2 J$ and $\delta V \approx 0.3$ J, assuming the inter-layer distance $d=0.7$ nm.  However, for other moir\'e bilayer with larger distance, like $d=2.5$ nm,  both $\delta J$ and $\delta V$ term can be larger than $J$. Especially $\delta V$ can be several times larger than $J$.

 To test the stability of the CSL, we include the $\delta V$ term and the $\delta J$ term in the iDMRG. The chirality order correlation is shown in Fig.~\ref{fig:dV} and we can see that the CSL survives at least up to $\delta V=0.5 J, \delta J=0.5 J$. Therefore we conclude that the CSL identified in the isotropic limit should be stable to anisotropic terms existing in twisted AB stacked TMD homo-bilayer with distance $d=0.7$ nm. For other moir\'e bilayer with larger $d$, a phase transition should happen towards a different phase when increasing $d$, which we leave to future study.

\section{Supersolids at imbalanced filling}
The previous part is focused on the balanced filling, where the two layers have the same density: $n_t=n_b=\frac{1}{2}$.  In this section we move to the imbalanced filling with $n_t=\frac{1}{2}(1+\delta)$ and $n_b=\frac{1}{2}(1-\delta)$. This corresponds to $P_z=\frac{1}{2}(n_t-n_b)=\delta$.  For the moir\'e bilayer, as the charges of the two layers are separately conserved, it is easy to tune $\delta$ from $0$ to $100\%$.  Here we show evidences for two different exciton condensation phases with inter-layer coherence at the $\delta \rightarrow 0$ and the $\delta \rightarrow 1$ limit respectively.

\subsection{Supersolid at the $\delta \rightarrow 0$ limit}
\begin{figure}[ht]
\centering
\includegraphics[width=0.98\textwidth]{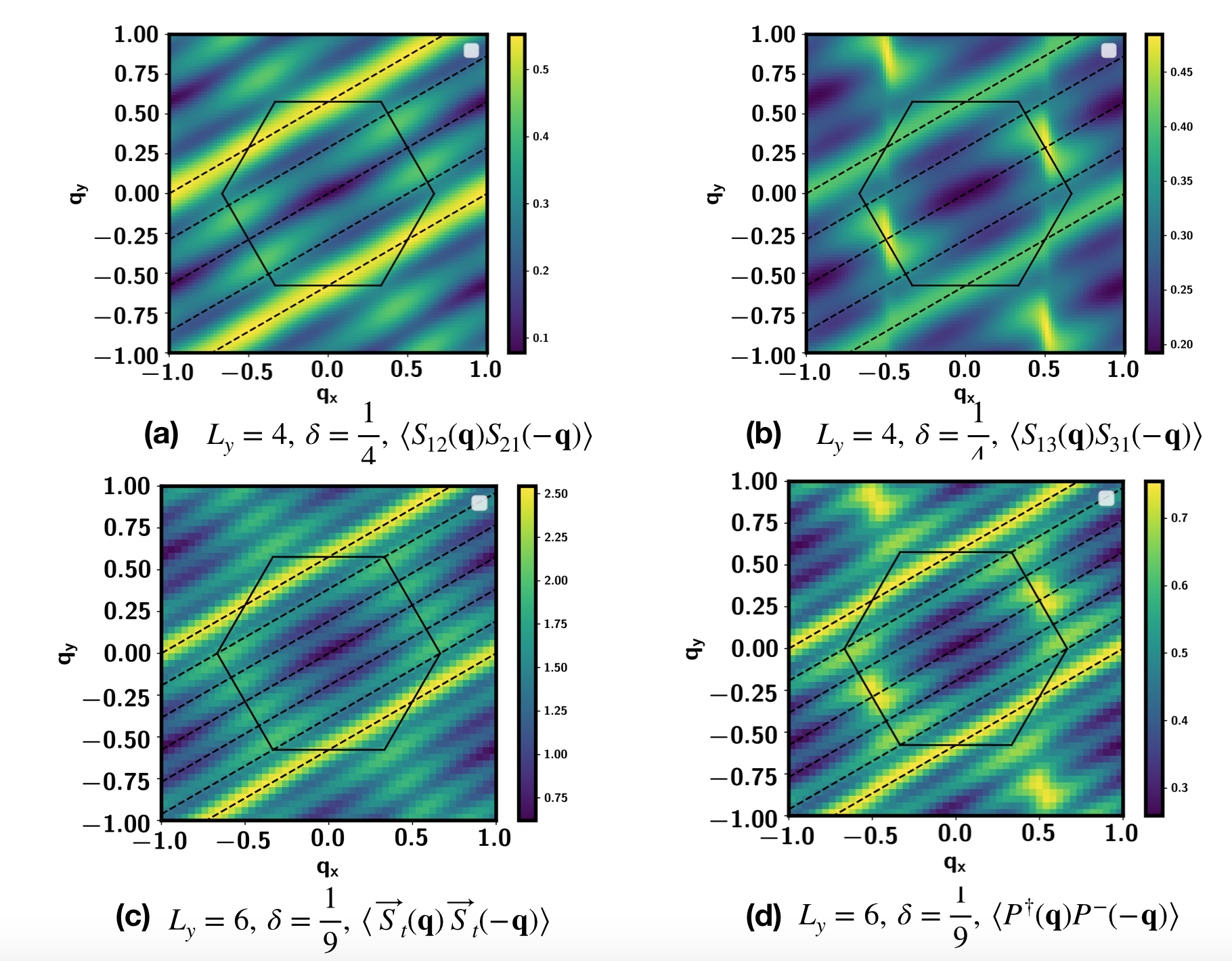}
\caption{Spin-spin structure factor at small $\delta$ for $L_y=4$ and $L_y=6$, obtained from infinite DMRG. Here we parameterize $\mathbf q=q_1 \mathbf b_1+q_2 \mathbf b_2$. $\mathbf b_1$ and $\mathbf b_2$ are two reciprocal vectors defined as $\mathbf a_i \cdot \mathbf b_j=2\pi \delta_{ij}$. The dashed lines are along the cut with $q_2=\frac{2\pi}{L_y} n$ with $n$ as an integer. $\vec{S_t}$ is the spin operator projected to the top layer. $\vec{S}_t(\mathbf q) \cdot \vec{S}_t(-\mathbf q) \sim S_{12}(\mathbf q) S_{21}(-\mathbf q)$. The structure factor for $S_b$ at the bottom layer is similar to that of $S_t$. $P^\dagger(\mathbf q) P^{-}(-\mathbf q) \sim S_{13}(\mathbf q) S_{31}(-\mathbf q)$ measures the correlation function of the exciton order parameter, which shows a peak at the M point with momentum $\frac{1}{2} \mathbf{b_1}$, indicating exciton condensation at non-zero momentum.}
\label{fig:SS_delta_1}
\end{figure}

\begin{figure}[ht]
\centering
\includegraphics[width=0.98\textwidth]{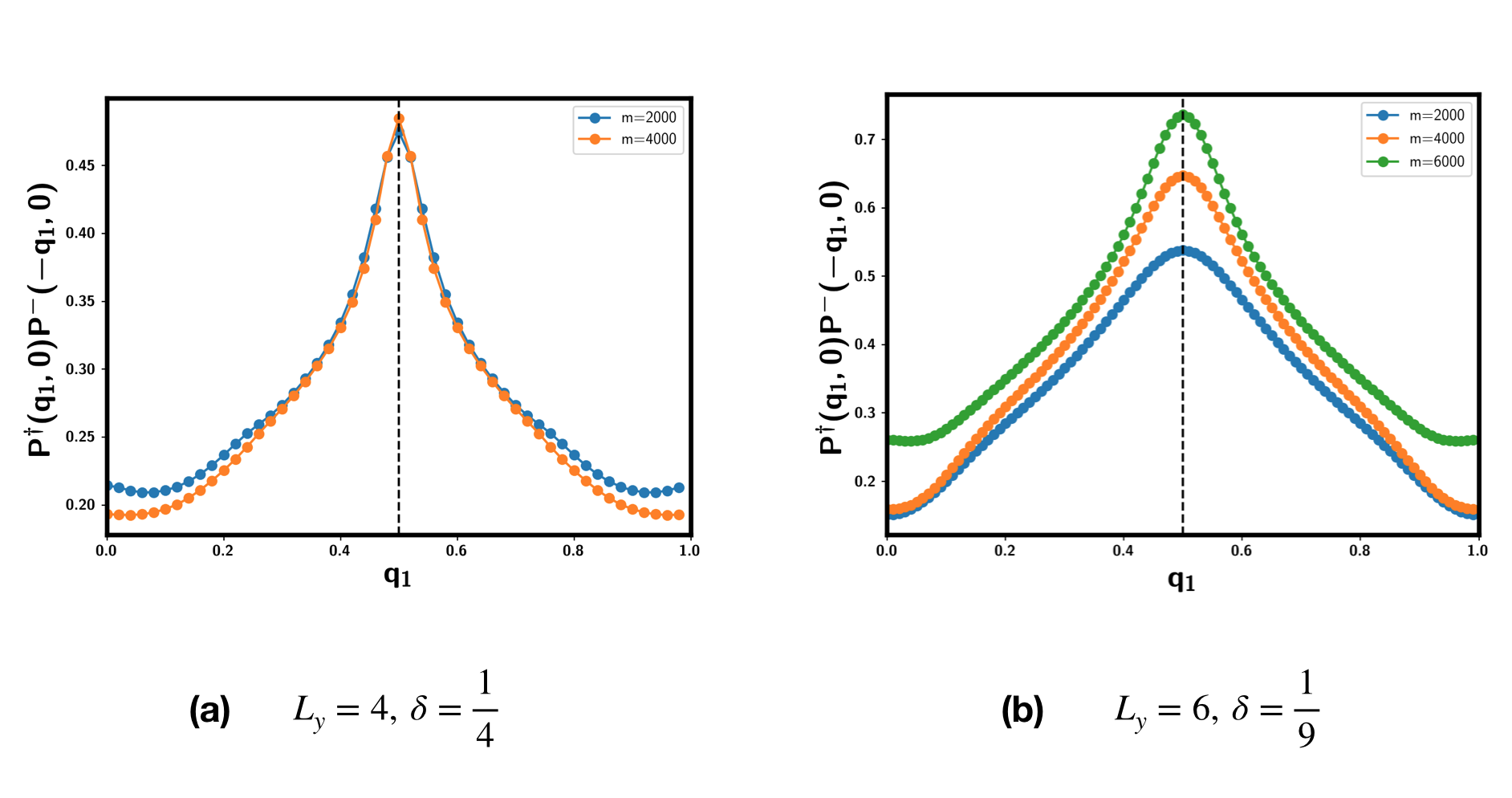}
\caption{Structure factor $P^\dagger(\mathbf q) P^-(-\mathbf q)$ along the cut $q_2=0$ for small $\delta$.  $P^\dagger(\mathbf q)P^{-}(-\mathbf q)\sim S_{13}(\mathbf q)S_{31}(-\mathbf q)$. $q_1$ is in unit of $|\mathbf b_1|$. The peak is at the $M$ point with momentum $\frac{1}{2} \mathbf{b_1}$, as denoted by the dashed lines.  The peak for $L_y=6$ is broad due to small correlation length limited by the bond dimension, but the peak grows sharper when increasing the bond dimension $m$.}
\label{fig:PP_cut_delta_1}
\end{figure}

At finite but small $\delta$, the system still shows a $2\times 1$ stripe structure in the bond order $\langle P_{ij}\rangle$ . For spin-spin correlation, $\vec{S}_t(x) \vec{S}_t(y)$ is short ranged. However, the exciton order $P^\dagger(x) P^{-}(y)$ is longer ranged with a momentum at $M$ point.   We parameterize $\mathbf q=q_1 \mathbf b_1+q_2 \mathbf b_2$. $\mathbf b_1$ and $\mathbf b_2$ are two reciprocal vectors defined as $\mathbf a_i \cdot \mathbf b_j=2\pi \delta_{ij}$. As shown in Fig.~\ref{fig:SS_delta_1}, the intra-layer spin-spin correlation (for example, $\vec{S}_{t}$) has no feature along $q_1$ direction, but has peak along $q_2=\frac{1}{2}$.  This is consistent with the decoupled stripe phase at the $\delta=0$ point.  However, the exciton order $P^\dagger \sim S_{13}$ shows a peak at momentum $M=\frac{1}{2} \mathbf b_1$.  We believe the exciton order is gapless and its correlation length grows with the bond dimension, as suggested by the plot in Fig.~\ref{fig:PP_cut_delta_1}.

In conclusion, the small $\delta$ limit has exciton condensed at momentum $M$ on top of the stripe phase.  We could view it as a supersolid with inter-layer coherence.

\subsection{Superfluid at the $\delta \rightarrow 1$ limit}

\begin{figure}[ht]
\centering
\includegraphics[width=0.98\textwidth]{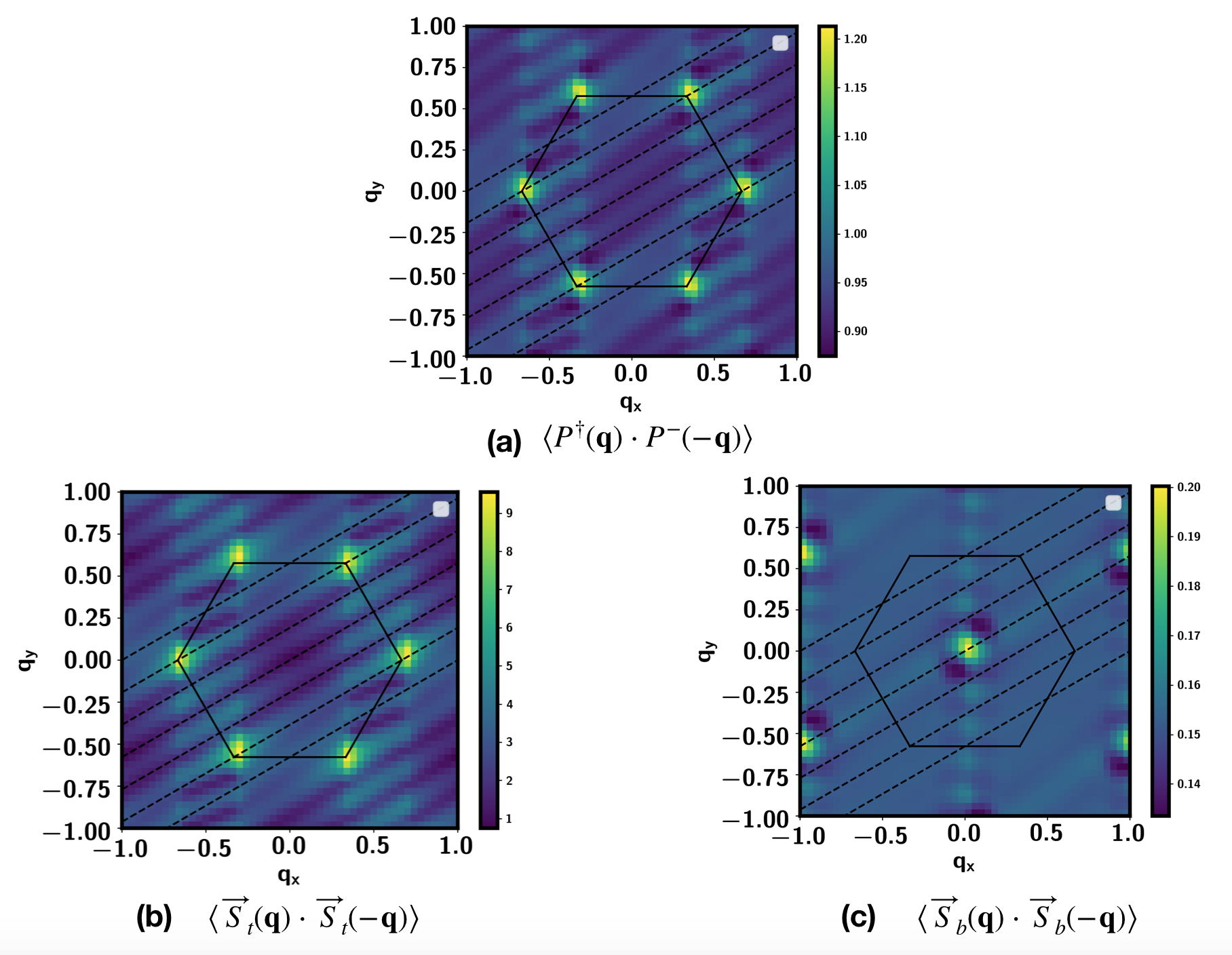}
\caption{Structure factor at $\delta=\frac{8}{9}$ for $L_y=6$. (a)Exciton order $\langle S_{13}(\mathbf q) S_{31}(-\mathbf q)$;  (b) Intra-layer spin spin correlation at top layer $S_{12}(\mathbf q) S_{21}(-\mathbf q)$; (c) Intra-layer spin-spin correlation at the bottom layer: $S_{34}(\mathbf q)S_{43}(-\mathbf q)$. }
\label{fig:SS_delta_8}
\end{figure}

We focus on the regime with $\delta=1-2x$ in the $x\rightarrow 0$ limit. At $x=0$ limit, the layer pseudospin is fully polarized and the model reduced to spin $1/2$ model on the top layer. The ground state is well known to be the $120^\circ$ magnetically ordered phase. At small $x$, we need to change the density of the top layer and the bottom layer to be $n_t=1-x$ and $n_b=x$, so there will be inter-layer excitons with density $x$. 

In the following we derive an effective model for the excitons at small $x$ limit. First, it is convenient to represent the electron operator in the top layer as:

\begin{equation}
  c_{i;t \sigma}=h_i^\dagger b_{i;\sigma}
\end{equation}
where $h_i^\dagger$ is a slave fermion operator which creates a hole for the top layer at site $i$ and $b_{i;\sigma}$ is the usual Schwinger boson for the top layer. We have the constraint $h^\dagger_i h_i+\sum_\sigma b^\dagger_{i;\sigma}b_{i;\sigma}=1$ at each site $i$. When $\delta=1-2x$, we have $\langle h^\dagger_i h_i \rangle=\sum_{\sigma}\langle c^\dagger_{i;b\sigma} c_{i;b\sigma}\rangle=x$ and $\sum_\sigma \langle b^\dagger_{i;\sigma}b_{i;\sigma}\rangle=1-x$.

\begin{figure}[ht]
\centering
\includegraphics[width=0.98\textwidth]{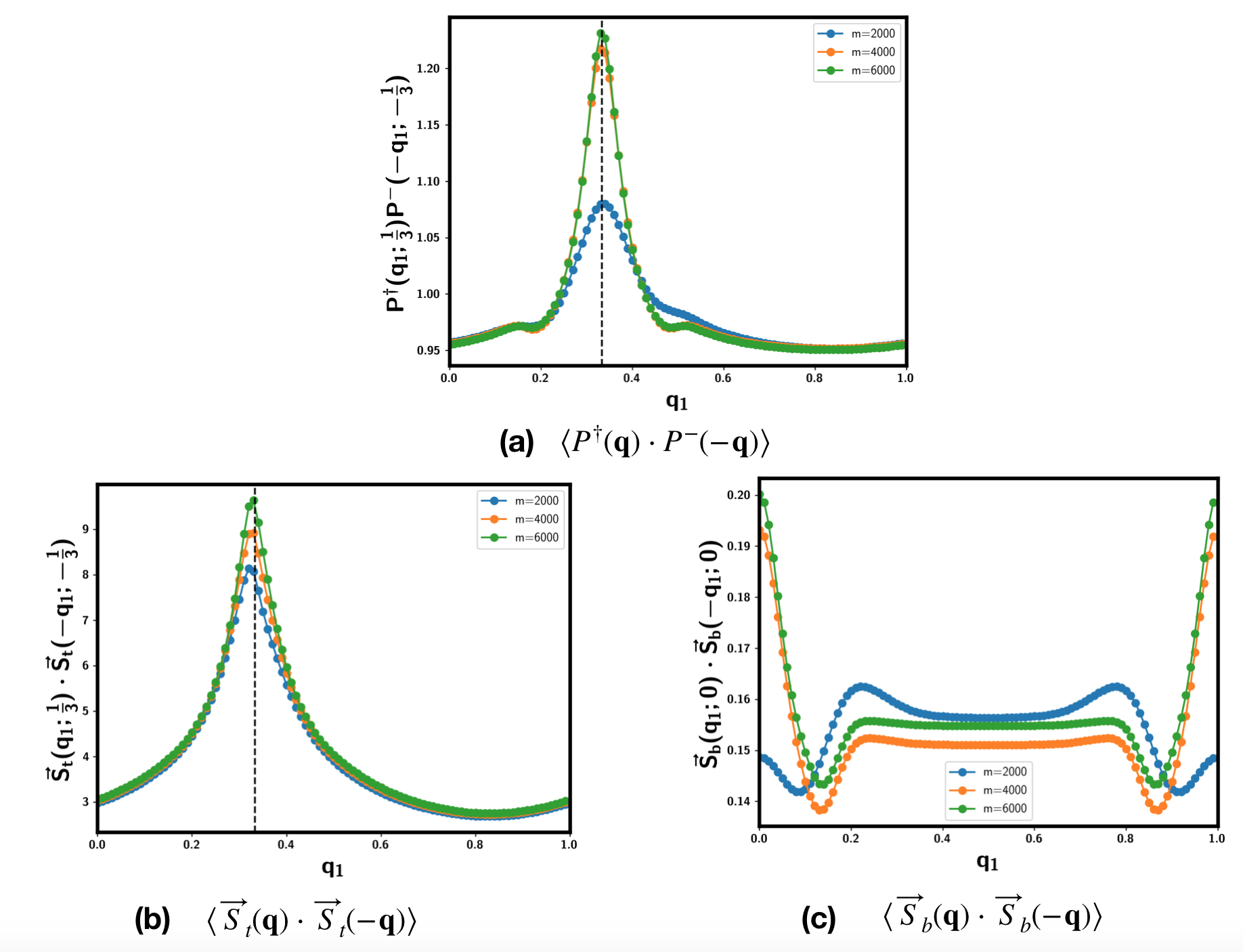}
\caption{Structure factor at $\delta=\frac{8}{9}$ along certain cuts for $L_y=6$.  }
\label{fig:PP_delta_8}
\end{figure}

Next we rewrite the super-exchange interaction as:

\begin{align}
  H_S&=-J \sum_{\langle ij \rangle} \sum_{\sigma,\sigma'=\uparrow,\downarrow}c^\dagger_{i;b \sigma}c_{j;b\sigma} c^\dagger_{j;t;\sigma'}c_{i;t;\sigma'}\notag+h.c.\\
&=-J\sum_{\langle ij \rangle}\sum_{\sigma}c^\dagger_{i;b \sigma}c_{j;b\sigma}h_j h^\dagger_i \sum_{\sigma'}b^\dagger_{j;\sigma'}b_{i;\sigma'}+h.c.
\end{align}
where we have ignored the terms with only intra-layer hopping proceses, which do not influence the dynamics of the excitons at the dilute limit $x\rightarrow 0$. 

Note that $b_\sigma$ represents the spin degree of freedom in the top layer. At the small $x$ limit, we know that the spin of the top layer is in the $120^\circ$ order. Thus we assume that the Schwinger boson condenses with expectation value

\begin{equation}
  \begin{pmatrix} b_{i;\uparrow} \\ b_{i;\downarrow} \end{pmatrix}=\sqrt{\frac{1-x}{2}} U_{i} \begin{pmatrix} 1 \\ 1 \end{pmatrix}
\end{equation}
where $U_i \in SU(2)$ rotates the spin  to the direction of the $120^\circ$ order with $A,B,C$ sublattice structure.  We have $U_A=-I$, $U_B=e^{i \frac{\pi}{3} \sigma_z}$ and $U_C=e^{-i \frac{\pi}{3}\sigma_z}$.   Here we add a minus sign to $U_A$ to make $\langle b^\dagger_{j;\sigma} b_{i;\sigma}\rangle=-\frac{1}{2} (1-x)$ for every bond.  A different gauge choice will break the $C_3$ symmetry for the hopping of the exciton, but it does not change the flux around one triangle. 

If we ignore the spin fluctuation (the goldstone modes) on top of the $120^\circ$ order, we can just substitute $b^\dagger_{j;\sigma}$ with its condensation expectation value, finally we get:

\begin{equation}
  H_S=\frac{1}{2}(1-x)J \sum_{ \langle ij \rangle }\sum_\sigma c^\dagger_{i;b \sigma}c_{j;b\sigma}h_j h_i^\dagger+h.c.
\end{equation}

Next we relabel $\Phi_{i;\sigma}=c_{i;b\sigma}h_i$, which creates an inter-layer exciton with spin $\sigma$ on the bottom layer. The exciton $\Phi_{i;\sigma}$ can be viewed as formed by combining electron in the bottom layer with a fermionic holon on the top layer. The dynamics of these excitons is governed by:

\begin{equation}
  H_S=-\frac{1}{2}(1-x)J \sum_{ \langle ij \rangle }\sum_\sigma \Phi^\dagger_{i;\sigma}\Phi_{j;\sigma}+h.c.
\end{equation}

So eventually we have a spin $1/2$ boson gas with total density $x$ per site on triangular lattice with an unfrustrated hopping $t_\Phi>0$. The ground state is known to be a spin polarized uniform BEC if the spin-spin coupling of the bottom  layer can be ignored at the small $x$ limit. $\Phi_{i;\sigma}$ carries dipole charge $P_z=1$ and thus there is a superfluid in counterflow transport.

We can verify the inter-layer exciton condensation picture  by calculating three different correlation functions:  (I) First, $\vec{S}_t$ should have the $120^\circ $ order as shown in Fig.~\ref{fig:SS_delta_8}(b); (II) $\vec{S}_b$ should have a FM order, as the spin of the bottom layer should be polarized, as is confirmed in Fig.~\ref{fig:SS_delta_8}(c)  (III) For $P^\dagger(\mathbf q) P^-(-\mathbf q)$, we note that $P^\dagger \sim c^\dagger_t c_b\sim b^\dagger h c_b\sim b^\dagger \Phi$.  Because $b$ condensed with momentum $\mathbf K, \mathbf K'$, $\Phi$ condenses with momentum $0$, $P^\dagger$ condenses with momentum $\mathbf K,\mathbf K'$.  We indeed confirm this in Fig.~\ref{fig:SS_delta_8}(a).  The peaks for these three correlation functions at the associated momentum all grow with the band dimension, as shown in Fig.~\ref{fig:PP_delta_8}.

\subsection{Expressions of correlation functions}

Some useful equations:

\begin{equation}
  4\vec{S}_{i;t}\cdot \vec{S}_{j;t}=S_{11}(i)S_{11}(j)+S_{22}(i)S_{22}(j)-S_{11}(i)S_{22}(j)-S_{22}(i)S_{11}(j)+2 S_{12}(i)S_{21}(j)+2S_{21}(i)S_{12}(j)
\end{equation}

\begin{equation}
  4\vec{S}_{i;b}\cdot \vec{S}_{j;b}=S_{33}(i)S_{33}(j)+S_{44}(i)S_{44}(j)-S_{33}(i)S_{44}(j)-S_{44}(i)S_{33}(j)+2 S_{34}(i)S_{43}(j)+2S_{43}(i)S_{34}(j)
\end{equation}

\begin{equation}
  4\vec{S}_{i;t}\cdot \vec{S}_{j;b}=S_{11}(i)S_{33}(j)+S_{22}(i)S_{44}(j)-S_{11}(i)S_{44}(j)-S_{22}(i)S_{33}(j)+2 S_{12}(i)S_{43}(j)+2S_{21}(i)S_{34}(j)
\end{equation}

\begin{equation}
  4\vec{S}_{i;b}\cdot \vec{S}_{j;t}=S_{33}(i)S_{11}(j)+S_{44}(i)S_{22}(j)-S_{33}(i)S_{22}(j)-S_{44}(i)S_{11}(j)+2 S_{34}(i)S_{21}(j)+2S_{43}(i)S_{12}(j)
\end{equation}

\begin{equation}
  P^\dagger_i   P^{-}_j=S_{13}(i)S_{31}(j)+S_{24}(i)S_{42}(j)+S_{24}(i)S_{31}(j)+S_{13}(i)S_{42}(j)
\end{equation}

\begin{equation}
  P_{i;z}   P_{j;z}=(S_{11}(i)+S_{22}(i)-S_{33}(i)-S_{44}(i))(S_{11}(j)+S_{22}(j)-S_{33}(j)-S_{44}(j))
\end{equation}

\end{document}